\documentclass[aps,prl,reprint,superscriptaddress,footinbib,showpacs,floatfix]{revtex4-2}

\usepackage{comment}
\usepackage{graphicx}
\usepackage{latexsym}
\usepackage{xcolor}
\usepackage{amsmath}
\usepackage{multirow}
\usepackage{booktabs}
\usepackage{textcomp}
\usepackage{placeins}
\usepackage{braket}
\usepackage{float}
\usepackage{ulem}
\usepackage[makeroom]{cancel}

\usepackage{hyperref}
\definecolor{linkblue}{RGB}{45,48,146}
\hypersetup{
  colorlinks=true,
  linkcolor=linkblue,
  citecolor=linkblue,
  urlcolor=linkblue,
}

\usepackage{bbm}
\usepackage{bm}
\usepackage{lipsum}

\usepackage{amsfonts,amssymb}

\newcommand{\Eqref}[1] {Eq.~\eqref{#1}}

\newcommand{\lk}	[0] {\mathrm{Lk}}

\newcommand{\lplec} [0] {L_\mathrm{plec}}
\newcommand{\lcond} [0] {L_\mathrm{cond}}

\newcommand{\sig} [0] {\sigma}
\newcommand{\sigs} [0] {\sigma_s}
\newcommand{\sigp} [0] {\sigma_p}

\newcommand{\ffe}[0]{F}
\newcommand{\cfe}[0]{G}

\newcommand{\sfed}[0]{{\cal S}}
\newcommand{\pfed}[0]{{\cal P}}

\begin{document}
\newcommand{\basetitle}{Tunable topological droplets on supercoiled DNA under tension}

\author{Enrico Skoruppa}
\thanks{esk.phys@gmail.com}
\affiliation{
    School of Physics and Astronomy, University of Edinburgh, Peter Guthrie Tait Road, Edinburgh EH9 3FD, United Kingdom
}
\affiliation{
    Cluster of Excellence Physics of Life, TU Dresden, Arnoldstraße 18, 01307 Dresden, Germany
}
\author{Prithwidip Saha}
\affiliation{
    Institute for Physics, University of Augsburg, Universitätsstrasse 1, 86159 Augsburg, Germany
}
\author{Jan Brugu{\'e}s}
\affiliation{
    Cluster of Excellence Physics of Life, TU Dresden, Arnoldstraße 18, 01307 Dresden, Germany
}
\affiliation{
    Max Planck Institute for the Physics of Complex Systems, N{\"o}thnitzer Str. 38, 01187 Dresden, Germany
}
\affiliation{
    Max Planck Institute of Molecular Cell Biology and Genetics, Pfotenhauerstraße 108, 01307 Dresden, Germany
}
\affiliation{
    Center for Systems Biology Dresden, Pfotenhauerstraße 108, 01307 Dresden, Germany
}
\author{Jan Lipfert}
\affiliation{
    Institute for Physics, University of Augsburg, Universitätsstrasse 1, 86159 Augsburg, Germany
}
\author{Davide Marenduzzo}
\affiliation{
    School of Physics and Astronomy, University of Edinburgh, Peter Guthrie Tait Road, Edinburgh EH9 3FD, United Kingdom
}
\author{Helmut Schiessel}
\thanks{helmut.schiessel@tu-dresden.de}
\affiliation{
    Cluster of Excellence Physics of Life, TU Dresden, Arnoldstraße 18, 01307 Dresden, Germany
}
\affiliation{
    Institut f{\"u}r Theoretische Physik, TU Dresden, 01062 Dresden, Germany
}

\title{\basetitle}

\begin{abstract}
    
DNA--protein co-condensation is an important regulatory mechanism for gene expression and repair. For straight DNA under constant stretching forces, co-condensation is ``all-or-nothing'': above a critical force, condensates are prevented from nucleating, while below it the entire molecule collapses into a droplet.
Here we show by coarse-grained simulations and analytic theory that DNA supercoiling mechanically differentiates the condensation substrate: DNA plectonemes constitute force-insulated pockets on which condensates can nucleate even above the critical force.
The size of the resulting droplets is tunable through the applied force and linking number.
Surface tension drives the coarsening of multiple plectoneme-associated condensates into a single topological droplet.
Magnetic tweezers experiments with spermine reproduce both the force-dependent collapse and the supercoiling-dependent condensation.
\end{abstract}

\maketitle

Biomolecular condensates are membraneless compartments formed by liquid--liquid phase separation (LLPS)~\cite{brangwynne09,banani17,aierken2026}.
These condensates are ubiquitous across cells and have been implicated in a wide range of biological functions, including ribosome biogenesis in the nucleolus~\cite{feric16}, transcriptional control~\cite{boija18}, and heterochromatin formation~\cite{strom17heterochromatin}.
DNA--protein co-condensates are an important subclass of these condensates~\cite{Quail2021,larson17hp1} and have been implicated, for example, in gene regulation~\cite{morin22klf4,renger22cocond} and DNA repair~\cite{chappidi24parp1,kilic19repair}.
They form either through polymer-assisted condensation, where a polymer acts as a local chemical-potential trap that condenses proteins below their bulk saturation concentration~\cite{sommer22pac,rouches25prewetting,Mukherjee2026}, or through bridging-induced phase separation~\cite{brackley13bips,ryu21cohesin}, where protein binding bridges DNA segments well separated in sequence space.

\begin{figure}[t]
\centering
\includegraphics[width=8.6cm]{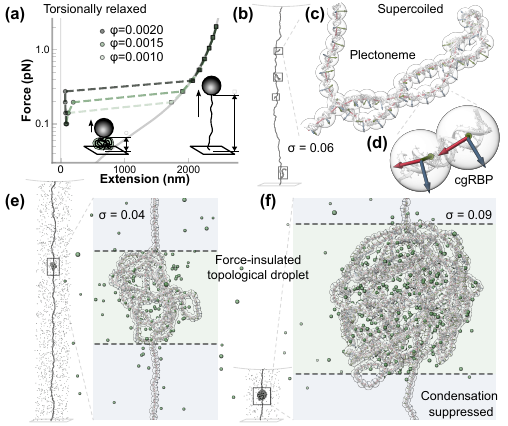}
\caption{
\textbf{All-or-nothing force gate and topological differentiation.}
(a)~Simulated force--extension curves of torsionally relaxed DNA tethers with condensing
proteins at volume fractions $\varphi = 0.0020$, $0.0015$, and $0.0010$; solid line: WLC prediction~\cite{Bouchiat1999}. Below a $\varphi$-dependent critical force $f_c$ the
entire tether collapses into a co-condensate.
(b)~Snapshot of a protein-free, torsionally constrained $8$\,kbp DNA tether
at supercoiling density $\sigma = 0.06$ and force $f = 2$~pN,
showing four coexisting plectonemic domains (boxes); close-up in (c).
(d)~Coarse-grained rigid-base-pair (cgRBP) representation used in the
simulations~\cite{lammps22,skoruppa26cgdna}.
(e,f)~Co-condensation on supercoiled tethers at $\sigma = 0.04$ (e) and
$\sigma = 0.09$ (f), both at $f = 2$~pN and $\varphi = 0.001$, i.e.,\ above
the critical force. The formation of plectonemes differentiates the binding
substrate, inducing a force-insulated pocket on which co-condensates can nucleate (green).
On the tension-bearing extended domains (blue), condensation remains suppressed. The size of the droplet
is tunable by the applied tension and supercoiling density.
}
\label{fig:fig1}
\end{figure}

Regulation by DNA--protein co-condensation requires control over droplet sizes and locations.
Spontaneous phase separation in free solution, however, leads to macrophase separation, where droplet size is only limited by the available material~\cite{berry18}.
In the dynamic and tightly organized nuclear environment, however, the molecular machinery exerts and responds to mechanical forces and torques: DNA is frequently held under tension~\cite{shin18condensates,kalukula22} and supercoiled by transcription and replication~\cite{liuwang87,naughton13}, and co-condensates themselves have been shown to be force-generating~\cite{Quail2021} and force-sensitive~\cite{nguyen22sox2}.
This suggests mechanical control parameters as natural candidates for the size regulation of DNA co-condensates.
Yet, at constant force and without explicit substrate differentiation, e.g., via sequence-dependent binding~\cite{morin22klf4}, the response is all-or-nothing: above a critical force, condensates are prevented from nucleating, while below it the entire molecule collapses into a droplet [Fig.~\ref{fig:fig1}(a)]~\cite{Quail2021,Haugk2024}.

Here we show, using coarse-grained molecular dynamics simulations [Fig.~\ref{fig:fig1}(d)]~\cite{skoruppa26cgdna} and an analytic two-phase theory, that DNA supercoiling, i.e., the over- or underwinding of the double helix characterized by the linking number $\lk$ of the two strands~\cite{bates05}, resolves the all-or-nothing response.
Torsionally constrained DNA partitions its excess linking number into twist and writhe~\cite{full71}. While tension suppresses writhe, sufficient supercoiling induces a buckling transition that splits the molecule into force-bearing extended domains and coiled plectonemic domains~\cite{mark95b,mark07,brac16,skor22} [Fig.~\ref{fig:fig1}(b,c)], the latter of which absorb most of the linking number added beyond buckling.
The spatial proximity of entry and exit points mechanically decouples plectonemes from the applied tension, turning plectonemic domains into effectively force-insulated pockets that differentiate the binding substrate, on which co-condensates can nucleate even above the critical force [Fig.~\ref{fig:fig1}(e,f)].
This is a passive, topological analogue of the tension shielding by active loop extrusion~\cite{takaki25}.
Importantly, as co-condensates are limited to plectonemic domains, whose extent is set by the imposed linking number and the applied force, DNA supercoiling allows for direct mechanical tuning of the size of the droplet formed.
Because of this topological size control mechanism, we refer to these condensates as ``topological droplets''.
Magnetic tweezers experiments with the polyamine spermine confirm both the force-dependent collapse and the supercoiling-dependent partial condensation [Fig.~\ref{fig:fig3}].

Without proteins, entropy favors fragmentation of the plectonemic phase of long tethers into multiple plectonemes~\cite{vanloenhout12,mark12,eman13,sege25}, since an extra plectoneme costs only a finite domain-wall energy. Co-condensing proteins introduce a three-dimensional dense phase with interfacial tension, so that the cost of splitting a droplet increases with its size. As we show below, entropy can then no longer sustain multiple domains, and condensates coarsen into a single droplet~[Fig.~\ref{fig:fig4}].

The coupling between protein condensation and supercoiling is potentially relevant to chromosome organization {\it in vivo}.
Several studies have demonstrated direct connections between DNA topology and condensation, highlighting that supercoiling can promote DNA condensation under tension~\cite{besteman07}, enhance ParB condensate formation~\cite{martingonzalez24parb}, modulate condensates formed through protein bridging~\cite{wei26corset}, lower the nucleation barrier for co-condensation on tension-free DNA loops~\cite{liu26supercoil}, and split annealed polyelectrolytes into coexisting twist- and writhe-rich domains~\cite{stano25}.
While these studies establish a connection between topology and condensation, the mechanisms by which DNA topology and mechanical tension control condensate size, and by which biomolecular condensates unify plectonemic domains into a single droplet, remain incompletely understood.

\textit{Simulation methods.}---We performed coarse-grained Langevin simulations of the DNA--protein system with LAMMPS~\cite{lammps22} (details in SI~\cite{SM}).
Tethers of $8$\,kbp DNA are modeled as twistable worm-like chains at $10$\,bp resolution with homogeneous bending and twisting stiffnesses~\cite{skor2025_cg,skoruppa26cgdna} and an effective excluded-volume diameter of $\approx 4$\,nm that accounts for electrostatic repulsion~\cite{Rybenkov1993,vand22} [see Fig.~\ref{fig:fig1}(d)].
Proteins are modeled as a Lennard--Jones fluid with attractive protein--protein and DNA--protein interactions.
As in magnetic tweezers~\cite{stri96,krie17a}, DNA is held under constant stretching force $f$ and at fixed supercoiling density $\sigma = \Delta\lk/\lk_0$, the excess linking number relative to relaxed DNA.
For the chosen protein--protein interaction (SI), proteins phase-separate on their own above the saturation volume fraction $\varphi_v\approx 0.00275$, located on the vapor branch of the liquid--vapor binodal~\cite{Vrabec2006}. All simulations were performed at protein volume fractions $\varphi<\varphi_v$, so that condensation can only occur on the DNA template.

\textit{Results.}---We first consider the force-extension response of torsionally relaxed tethers for various protein volume fractions $\varphi$ [Fig.~\ref{fig:fig1}(a)]. At large forces, the extension response matches the classical worm-like chain (WLC) result~\cite{bust94,mark95,Bouchiat1999}.
As the force is lowered below a $\varphi$-dependent critical force $f_c$, the extension collapses to near zero, marking the formation of a DNA--protein co-condensate that absorbs the entire tether~\cite{Quail2021,Haugk2024}.

This all-or-nothing response can be understood from the free energy balance of partitioning the molecule into condensed and extended domains. Extending the approach of Quail et al.~\cite{Quail2021} by a bending term, we write the free energy of a co-condensate containing DNA of contour length $\lcond$ as
\begin{equation}
    \label{eq:cfe}
    \cfe(\lcond)=-a\lcond+b\lcond^{2/3}+c\lcond^{1/3},
\end{equation}
with $a,b,c\geq0$. These terms describe the bulk free energy, the interfacial free energy, and the cost of bending a semi-flexible polymer into a droplet~\cite{Odijk1983,GrosbergKhokhlov1994,Sakaue2007}. As the last term is subdominant, we set $c=0$ in all fits (SI), which changes nothing qualitatively.
 Denoting by $-g(f)$ the free energy density of stretched DNA~\cite{mark95,mark07,skor22}, the total free energy of a molecule of contour length $L$ with fraction $\zeta$ absorbed into condensates reads $F(\zeta) = -(1-\zeta)L\,g(f) + \cfe(\zeta L)$. As $F$ is concave in $\zeta$, its minima must lie at the boundary values $\zeta=0$ or $\zeta=1$. Tension alone, therefore, admits no stable finite partition: below a critical force the condensate occupies the whole polymer, while above it no condensate forms.
The critical force $f_c$ is set by $L\,g(f_c)=-\cfe(L)$, and increases with protein volume fraction $\varphi$~[Fig.~\ref{fig:fig1}(a)].

\begin{figure}[t]
\centering
\includegraphics[width=8.6cm]{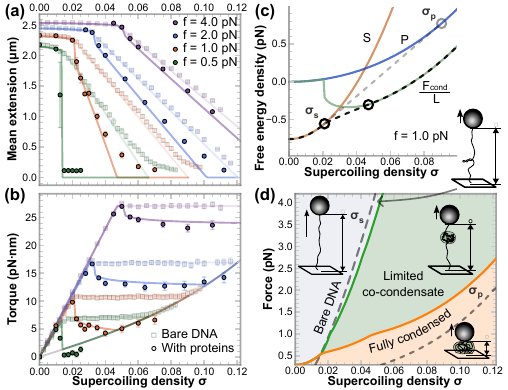}
\caption{
\textbf{Rotation curves and co-condensation phase diagram.}
(a,b)~Mean extension (a) and torque (b) of $8$\,kbp tethers versus supercoiling
density $\sig$ at $f = 0.5$, $1.0$, $2.0$, $4.0$\,pN. Squares:
protein-free simulations; circles: with proteins ($\varphi = 0.001$);
lines: classical (transparent) and co-condensate two-phase theory (saturated).
(c)~Free energy densities of the stretched ($\sfed$) and plectonemic ($\pfed$) phases, and of plectonemic DNA with co-condensate ($\ffe_{\mathrm{cond}}/L$), at $f = 1.0$\,pN.
The gray dashed line is the protein-free common tangent. The black dashed line is the minimized free energy density of the co-condensate model. Circles mark the onset of co-condensation and full absorption (black: with condensate; gray: bare DNA).
(d)~Phase diagram for topological droplets. The dashed lines mark $\sigs$ and $\sigp$ of the condensate-free state. The three shaded regions mark bare DNA tethers (blue), tethers with a limited co-condensate (green), and fully condensed tethers (orange).
}
\label{fig:fig2}
\end{figure}

Without proteins, simulated rotation curves (extension and torque vs $\sig$ at constant force) of torsionally constrained tethers [squares in Fig.~\ref{fig:fig2}(a,b); writhe in SI] reproduce the well-documented behavior seen in magnetic tweezers experiments~\cite{stri96,moro98,mark07,lipf10,vand22,skor22,sege25}: at low $\sig$, most of the linking number takes the form of twist, leading to a linear torque response and a modest quadratic decrease of the extension; beyond a buckling point $\sigs$ that increases with force, most of the additional linking number is absorbed as writhe, leading to the formation of plectonemes, which take up a growing share of the contour length [Fig.~\ref{fig:fig1}(b,c)]. As twist stays approximately constant in this regime, torque saturates on a post-buckling plateau~\cite{mark07,fort08,sege25}.
Since the entry and exit points of plectonemes nearly coincide, the absorbed contour length does not contribute to the extension, which therefore decreases linearly with $\sig$, with a slope that becomes shallower as the force increases~\cite{mark07,skor22}.

The introduction of proteins changes this picture [circles in Fig.~\ref{fig:fig2}(a,b); $\varphi=0.001\approx0.36\,\varphi_v$, further $\varphi$ and DNA--protein attractions in SI]. Above the critical force [$f_c\approx 0.18$\,pN at this $\varphi$; Fig.~\ref{fig:fig1}(a)], co-condensates are prevented from nucleating on the tension-bearing extended DNA, but can nucleate on the force-insulated plectonemes [Fig.~\ref{fig:fig1}(e,f)].
Once a condensate forms, the extension drops abruptly and then decreases more steeply with $\sig$. The torque shows a similar drop, before settling on a lower post-buckling plateau. This suggests that co-condensed plectonemic domains experience an effectively reduced force, consistent with the findings of Quail et al.~\cite{Quail2021}, and with wetting-enhanced twist-to-writhe conversion in torsionally constrained loops~\cite{liu26supercoil}. The change in response becomes more pronounced as the force decreases, until at $0.5$\,pN the tether collapses fully at finite $\sig$. These observations point to a condensate-induced shift of the two-phase equilibrium.

These observations can be captured by extending the two-phase model of supercoiled DNA~\cite{mark95b,mark07}. It expresses the free energy of a tether of length $L$ with plectonemic fraction $\chi$ as $\ffe=(1-\chi)L\,\sfed(\phi,f)+\chi L\,\pfed(\psi)$, where $\phi$ and $\psi$ are the supercoiling densities in the stretched and plectonemic domains and $\sfed$ and $\pfed$ the corresponding free energy densities [Fig.~\ref{fig:fig2}(c); explicit forms in SI]. Minimization at fixed
$\sig$
yields a common-tangent construction [gray in Fig.~\ref{fig:fig2}(c)] with coexistence between the buckling point $\sigs$ and $\sigp$, where the tether becomes fully plectonemic, and reproduces the protein-free rotation curves [transparent lines in Fig.~\ref{fig:fig2}(a,b)]. To include co-condensation, we incorporate the co-condensate free energy [\Eqref{eq:cfe}] into the plectonemic term
\begin{equation}
    \label{eq:Fcond}
    F_{\mathrm{cond}} = \lplec\,\pfed(\psi) + \cfe(\lplec)\,\Theta\left(-\cfe(\lplec)\right),
\end{equation}
with $\lplec=\chi L$. Here the Heaviside step function $\Theta$ admits a droplet only where it lowers the free energy.
The condensate term modifies the coexistence construction [black in Fig.~\ref{fig:fig2}(c)], and fitting $a$ and $b$ (separate values at $0.5$\,pN; SI) accounts for the simulated curves with proteins [saturated lines in Fig.~\ref{fig:fig2}(a,b)]. The sub-extensive interfacial term causes a rebalancing of torque across the coexistence region, which relaxes as the droplet grows, and forbids droplets below a minimum size, causing the jumps in extension and torque. These jumps are akin to the torque ``overshoot'' at the buckling transition in short tethers~\cite{brutzer10,mark12}.
For large droplets, the bulk term dominates. It lowers the plectoneme free energy, and thereby allows plectonemic domains to absorb more contour length and linking number, which lowers the torque and steepens the extension slope.

The extended model yields the phase diagram of Fig.~\ref{fig:fig2}(d) (SI; at forces other than the simulated ones, $a$ and $b$ are linearly interpolated or extrapolated). At low $\sig$, the model predicts the all-or-nothing response of torsionally unconstrained DNA [Fig.~\ref{fig:fig1}(a)]: the tether transitions directly from extended to fully condensed. Above a threshold $\sig$, this boundary bifurcates into the onset of co-condensation (green line) and full absorption (orange line), opening a region in which bare extended DNA coexists with a co-condensed plectonemic domain. While $\sigp$ lies inside the fully condensed region, the condensate-free buckling line $\sigs$ can lie in any of the three regions. When it lies in the bare region, buckling precedes co-condensation. In contrast, when it lies in either co-condensate region, buckling shifts to the onset of co-condensation (green line) and the droplet forms together with the plectonemes. This behavior highlights the two-way feedback between plectonemes and co-condensates.
The phase diagram thus captures the central distinction between torsionally unconstrained and constrained DNA: topology replaces the all-or-nothing force gate with a finite regime of topological droplets.

\begin{figure}[t]
\centering
\includegraphics[width=8.6cm]{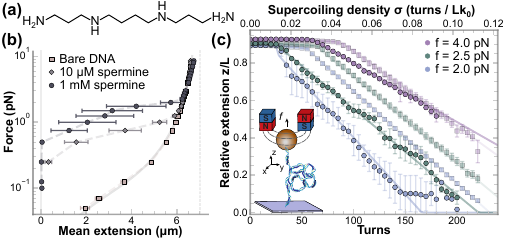}
\caption{
\textbf{Magnetic tweezers experiments with spermine.}
(a)~Chemical structure of the tetravalent polyamine spermine.
(b)~Force--extension curves of $20.6$\,kbp DNA without spermine (squares; gray line: WLC) and with $10$\,µM (diamonds) and $1$\,mM spermine (circles). Below a concentration-dependent critical force, the tether collapses.
(c)~Rotation curves at $f=2$--$4$\,pN without (squares) and with $1$\,mM spermine (circles).
}
\label{fig:fig3}
\end{figure}

To test our predictions experimentally, we performed magnetic tweezers (MT) measurements~\cite{krie17a}. We used a custom-built single-molecule MT setup with $20.6$\,kbp DNA~\cite{lipf09b,lipf10b} tethered between a flow cell surface and superparamagnetic beads~\cite{walk18,vand22,stor26,stor25} (further details in SI). In the MT, the applied force and linking number are controlled by translating and rotating the magnets~\cite{lipf09a}. Instead of a protein with potentially specific and directional interactions, we used the biologically relevant DNA condensing agent spermine~\cite{gosu76,besteman07,todd08} [Fig.~\ref{fig:fig3}(a), details in SI].
As the applied force was lowered stepwise in the presence of spermine, we observed a sudden collapse of the DNA [Fig.~\ref{fig:fig3}(b); time traces in SI], similar to the behavior predicted from simulations [Fig.~\ref{fig:fig1}(a)]. The critical force at which collapse occurs increases with increasing spermine concentration, in agreement with the simulations and with previous measurements~\cite{besteman07,todd08}. At fixed spermine concentration and constant force, rotation curves [Fig.~\ref{fig:fig3}(c)] again closely resemble the simulated ones [Fig.~\ref{fig:fig2}(a)].
Above the critical force for relaxed DNA, we observed a marked decrease in extension---without full collapse---compared to the spermine-free case, consistent with our predicted response for partial co-condensates.
Besteman et al.\ showed that imposed twist promotes condensation, which nucleates at a plectoneme~\cite{besteman07}; the partial condensation we observe above the critical force goes beyond these findings. We note that spermine deviates in molecular details from our simulated model protein, in particular due to strong electrostatic repulsion, non-ideal solution behavior at higher concentrations~\cite{todd08}, and specific binding modes to the DNA grooves. Even without condensation, polyamines alter the bending stiffness, the twist--writhe balance of DNA~\cite{baum97,shao12}, and the coiling density of plectonemes. Nonetheless, our experiments confirm both the force-dependent collapse and our prediction of supercoiling-dependent partial condensation.

\begin{figure}[t]
\centering
\includegraphics[width=8.6cm]{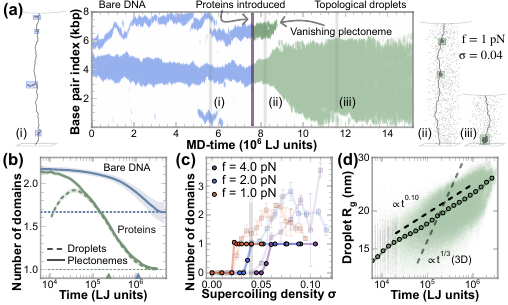}
\caption{
\textbf{Droplet-induced plectoneme coarsening.}
(a)~Kymograph of plectonemic domains at $f=1$\,pN and $\sig=0.04$ ($8$\,kbp). Without proteins, several plectonemes coexist (i); after protein addition (purple line), two droplets form (ii) and then ripen into a single one (iii).
(b)~Mean number of plectonemes (solid) and droplets (dashed) versus time after protein addition (green), and of plectonemes without proteins (blue), under the conditions of (a). Dotted: steady-state values; triangles: median time to lose the first plectoneme.
(c)~Mean steady-state number of droplets (circles) and of protein-free plectonemes (squares) versus $\sig$ [simulations of Fig.~\ref{fig:fig2}(a,b)]; gray bar: conditions of (a).
(d)~Radius of gyration of the surviving droplet versus time after protein addition; circles: binned mean; dashed: fit $R_g\propto t^{0.10}$.
}

\label{fig:fig4}
\end{figure}

Our simulations allow us to directly monitor the kinetics of topological droplet formation~[Fig.~\ref{fig:fig4}], which is difficult to study experimentally.
Simulations in the absence of proteins [Fig.~\ref{fig:fig4}(a), before protein addition] reveal a dynamic multiplectoneme configuration~\cite{eman13,sege25}, which is entropically favored, as expected for a one-dimensional system with only short-ranged interactions~\cite{huang2008}.
Introducing proteins [Fig.~\ref{fig:fig4}(a), after protein addition] drives coarsening into a single topological droplet.

Without proteins, the effectively 1D phase separation between plectonemic and stretched domains~\cite{mark07} is arrested by noise~\cite{forte2019}. Starting from configurations with at least two plectonemes, the mean number of plectonemes relaxes to a steady-state value of $\approx 1.7$ [Fig.~\ref{fig:fig4}(b)].
With proteins, the interfacial free energy of the three-dimensional condensate lifts this arrest and also speeds up coarsening: the first plectoneme is lost about five times faster than without proteins [triangles in Fig.~\ref{fig:fig4}(b)].
This behavior is robust: over the whole range of $\sigma$ and $f$ in which droplets form, a single droplet is stable in steady state, whereas protein-free tethers carry multiple plectonemes [Fig.~\ref{fig:fig4}(c)].

As for Cahn--Hilliard growth in model B~\cite{chaikin1995}, the fusion of small droplets into a single large droplet can occur through diffusion and coalescence of smaller droplets.
However, this process typically proceeds quite slowly; a faster mechanism is an analogue of Ostwald ripening~\cite{bray1994}, in which larger droplets grow at the expense of smaller ones due to differences in Laplace pressure~\cite{vorhees1992}. In our simulations, $\approx 89\%$ of the smaller droplets vanish by ripening (see SI).
For Cahn--Hilliard growth in 3 dimensions, we therefore expect the radius of a coarsening droplet to increase as $t^{1/3}$~\cite{chaikin1995}.
Instead, the radius of gyration of the surviving droplet grows only as $R_g\propto t^{0.10}$ [Fig.~\ref{fig:fig4}(d)], close to $t^{1/9}$.
This suggests that the growth of topological droplets is dominated by the 1D coarsening of plectonemes, for which domains grow as $t^{1/3}$~\cite{forte2019}: because the plectonemic length scale is a contour length along the polymer, the radius of the topological droplet scales with this length scale to the power $1/3$, giving $t^{1/9}$ (see SI).
In other words, our system shares 3D-like and 1D-like features: surface tension works in 3D, while the dynamics are dominated by 1D processes. Therefore, the dynamics of topological droplets are fundamentally different from those of liquid--liquid and bridging-induced phase separation~\cite{berry2015,brackley2017}, and the power law associated with their formation provides a simple signature of this distinction.

\textit{Conclusions.}---In summary, we have shown that coupling DNA supercoiling to co-condensation
leads to the formation of topological droplets and changes the nature of the phase separation along the molecule. On its own, supercoiling produces a two-fluid mixture of stretched and plectonemic DNA, for which the effectively 1D, short-ranged interactions lead to an Ising-like multiplectoneme state rather than a conventional phase transition. Introducing proteins that bind to DNA and to one another instead creates a 3D condensed phase with a finite surface tension. This interfacial cost drives coarsening into
a single macroscopic topological droplet, in line with the Peierls argument for Ising models~\cite{huang2008}, by which interfacial costs that grow with domain size enable phase ordering in more than one dimension.
Unlike condensation on a torsionally unconstrained polymer, the size of the 3D topological droplet is set by the amount of plectonemic DNA and can therefore be tuned mechanically through supercoiling and applied force. Magnetic tweezers experiments with spermine confirm the force-dependent collapse and the supercoiling-dependent partial condensation.

Topological droplets provide a simple physical mechanism for forming and stabilizing finite supercoiling domains. By concentrating proteins on plectonemic DNA and, through their surface tension, counteracting the fragmentation of plectonemes along the polymer, such condensates could localize on the chain and slow supercoiling diffusion. This mechanism is potentially relevant to the formation of topological domains in bacterial DNA~\cite{sinden1981}, which is both supercoiled and rich in genome-binding proteins. As transcription and replication generate supercoiling~\cite{liuwang87}, this mechanism also provides a possible feedback between topological stress and protein recruitment, in which supercoiling promotes condensation that in turn stabilizes the local topological state.

\textit{Acknowledgments.}---This work was supported by the Deutsche Forschungsgemeinschaft (DFG, German Research Foundation) under Germany’s Excellence Strategy (Grant No. EXC-2068-390729961) and through Research Training Group RTG 3120 ``Biomolecular Condensates'' (Project ID 542285965), by the European Union’s Horizon Europe research and innovation programme under the Marie Skłodowska-Curie Grant Agreement No. 101202019, by Grant No. NSF PHY-2309135 to the Kavli Institute for Theoretical Physics (KITP), the Gordon and Betty Moore Foundation Grant No. 2919.02, by the European Research Council Consolidator Grant ``ProForce'' (101002656), by the European Research Council Consolidator Grant ``SynthNuc'' (101045468), and by the Wellcome Trust (grant number 223097/Z/21/Z).
The authors gratefully acknowledge the computing time made available to them on the high-performance computer at the NHR Center of TU Dresden. This center is jointly supported by the Federal Ministry of Research, Technology and Space of Germany and the state governments participating in the NHR (www.nhr-verein.de/unsere-partner).

\bibliography{references}

\end{document}


\newcommand{\basetitle}{Tunable topological droplets on supercoiled DNA under tension}

\author{Enrico Skoruppa}
\thanks{esk.phys@gmail.com}
\affiliation{
    School of Physics and Astronomy, University of Edinburgh, Peter Guthrie Tait Road, Edinburgh EH9 3FD, United Kingdom
}
\affiliation{
    Cluster of Excellence Physics of Life, TU Dresden, Arnoldstraße 18, 01307 Dresden, Germany
}
\author{Prithwidip Saha}
\affiliation{
    Institute for Physics, University of Augsburg, Universitätsstrasse 1, 86159 Augsburg, Germany
}
\author{Jan Brugu{\'e}s}
\affiliation{
    Cluster of Excellence Physics of Life, TU Dresden, Arnoldstraße 18, 01307 Dresden, Germany
}
\affiliation{
    Max Planck Institute for the Physics of Complex Systems, N{\"o}thnitzer Str. 38, 01187 Dresden, Germany
}
\affiliation{
    Max Planck Institute of Molecular Cell Biology and Genetics, Pfotenhauerstraße 108, 01307 Dresden, Germany
}
\affiliation{
    Center for Systems Biology Dresden, Pfotenhauerstraße 108, 01307 Dresden, Germany
}
\author{Jan Lipfert}
\affiliation{
    Institute for Physics, University of Augsburg, Universitätsstrasse 1, 86159 Augsburg, Germany
}
\author{Davide Marenduzzo}
\affiliation{
    School of Physics and Astronomy, University of Edinburgh, Peter Guthrie Tait Road, Edinburgh EH9 3FD, United Kingdom
}
\author{Helmut Schiessel}
\thanks{helmut.schiessel@tu-dresden.de}
\affiliation{
    Cluster of Excellence Physics of Life, TU Dresden, Arnoldstraße 18, 01307 Dresden, Germany
}
\affiliation{
    Institut f{\"u}r Theoretische Physik, TU Dresden, 01062 Dresden, Germany
}

\title{Supporting Information for ``\basetitle''}

\maketitle

\section{Molecular dynamics simulations}
\label{sec:app_md}
\begin{figure*}[t]
\centering
\includegraphics[width=17.8cm]{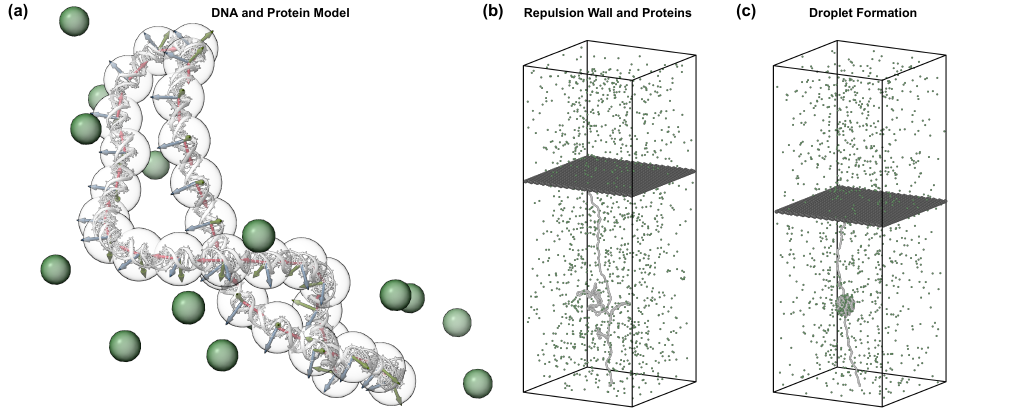}
\caption{
\textbf{Coarse-grained tweezer simulation of supercoiled DNA with co-condensing proteins.}
(a)~Representation of DNA and proteins. DNA is modeled as a chain of beads, each representing 10 base pairs and carrying a body-frame triad (arrows) that encodes the local bending and twist. The DNA is surrounded by spherical protein particles (green).
(b)~Simulation box. The DNA is tethered between a fixed bottom anchor and a rigid honeycomb wall (dark). The first bead is held fixed in space, with a repulsive wall preventing the chain from passing below it; the last bead is rigidly bound to the honeycomb wall, with which it forms a single rigid body that is pulled along $+z$ at constant force. The orientations of the wall and of the attached DNA bead are fixed. The wall spans the periodic $x$--$y$ cross-section, and because both ends are rotationally clamped, the linking number of the tethered molecule is conserved. The wall interacts repulsively with the DNA but is transparent to the proteins, which permeate it freely. Proteins (green) are dispersed throughout the box, which is periodic in $x$ and $y$ and bounded by reflective walls along the stretching axis $z$.
(c)~At sufficient attraction, the proteins nucleate a co-condensate droplet on the plectonemically coiled DNA substrate.
}
\label{fig:App_MD}
\end{figure*}

We study the interplay between DNA supercoiling and protein--DNA co-condensation with molecular-dynamics (MD) simulations of individual, end-tethered DNA molecules held under constant tension and imposed winding number in the presence of condensate-forming proteins, mimicking the setup of magnetic tweezers experiments (Fig.~\ref{fig:App_MD}). Both DNA and proteins are captured by coarse-grained representations, and all simulations are carried out with LAMMPS~\cite{lammps22} in reduced Lennard--Jones (LJ) units, in which energies are measured in units of $\kt$ (reduced temperature $T=1$) and lengths in units of the DNA bead spacing $\sigma$, i.e., the equilibrium distance between consecutive beads ($10\,\mathrm{bp} = 3.4\,\mathrm{nm}$; see below).
Following the convention for LJ systems, $\sigma$ and the pair parameters $\sigma_{\mathrm{x}}$ denote lengths in Secs.~\ref{sec:app_md}, \ref{sec:lj}, and \ref{sec:app_dropfinder}; everywhere else, $\sigma$ denotes the supercoiling density.

\subsection{DNA model}

DNA is represented as a coarse-grained version of the rigid base-pair (RBP) model, using the custom variable-resolution LAMMPS interaction styles introduced in Ref.~\cite{skoruppa26cgdna}. In this implementation, each monomer is a rigid body (element of $\SET$), represented as an ellipsoid in LAMMPS, that carries an intrinsic orientation encoded by a body-frame triad (Fig.~\ref{fig:App_MD}a). Rotational deformations encoding bending and twisting degrees of freedom between consecutive beads are parametrized as Euler vectors corresponding to the relative orientation of the respective body frames. The elastic energy is a quadratic form of the six relative $\SET$ coordinates (Euler vector components plus three translations). While the implementation supports sequence-dependent intrinsic shape and stiffness~\cite{skor2025_cg}, we use sequence-averaged, local elastic parameters with a straight ground state, i.e., a discretization of the twistable worm-like chain.

We use a discretization of $10\,\mathrm{bp}$ per bead and identify the resulting bead spacing with the unit length of the simulation, $\sigma = 3.4\,\mathrm{nm}$. The rotational couplings are set to reproduce a bending stiffness of $A = 42\,\mathrm{nm}$ and a torsional stiffness of $C = 86\,\mathrm{nm}$. Diagonal translational stiffnesses are set to $200\,\kt / \sigma^2$ each to obtain a near-inextensible chain.

Excluded volume interactions and electrostatic repulsion, which are necessary to prevent violations of the topological state and to impose reasonable coiling densities at large linking number excess, are jointly accounted for by repulsive LJ potentials (see below). Following Ref.~\cite{Rybenkov1993} and the fitting to experimental data in Ref.~\cite{vand22}, we set the effective DNA diameter to $1.2\,\sigma$ (i.e., $\approx 4 \, \mathrm{nm}$) to capture the effect of electrostatics at $150\,\mathrm{mM}$ monovalent salt.

Moreover, to prevent chain crossings via bond-overstretching and the associated loss of the topological state of the system, we supplement the RBP elasticity with a modified finitely extensible nonlinear elastic (FENE) term, acting on the center-to-center distance $r$ of bonded beads,
\begin{equation}
    \label{eq:fene}
    E_{\mathrm{FENE}}(r) = -\tfrac{1}{2}\,k_{\mathrm{eff}}\,
    \ln\!\left[1 - \frac{(r-R_c)^2}{(R_0-R_c)^2}\right]\,\Theta(r-R_c),
\end{equation}
with $k_{\mathrm{eff}} = K\,(R_0-R_c)^2$. The Heaviside step function $\Theta$ restricts the term to bond lengths beyond the threshold $R_c$, such that the harmonic elasticity alone governs the spectrum of moderate fluctuations. As usual, the FENE term diverges as $r \to R_0$. This FENE interaction is integrated into the bond style of the RBP model using the \texttt{rbpfene} bond style.
For the present simulations we use $K = 500$, $R_c = 1.15\,\sigma$, and $R_0 = 1.3\,\sigma$ (reduced units).

For additional details on the elastic energy and the force-field implementation, we refer to Ref.~\cite{skoruppa26cgdna}.

\subsection{Protein model and pairwise interactions}

Proteins are modeled as spherical particles of diameter $\sigma_{\mathrm{prot}} = 2\,\sigma$ ($\approx 6.8\,\mathrm{nm}$) with attractive LJ interactions to DNA and other proteins.
Simulations are conducted at fixed protein volume fraction
\begin{equation}
    \varphi = \frac{N_{\mathrm{prot}}\, V_{\mathrm{prot}}}{V_{\mathrm{box}}},
\end{equation}
with
\begin{equation}
    V_{\mathrm{prot}} = \frac{4}{3}\pi\left(\frac{\sigma_{\mathrm{prot}}}{2}\right)^{3},
\end{equation}
$N_{\mathrm{prot}}$ the number of proteins and $V_{\mathrm{box}}$ the volume of the simulation box.

All non-bonded pairwise interactions are modeled via truncated Lennard--Jones potentials,
\begin{equation}
    V_{\mathrm{LJ}}^{\mathrm{cut}}(r_{ij}) = \left[ V_{\mathrm{LJ}}(r_{ij}) - V_{\mathrm{LJ}}(r_{c,\mathrm{x}}) \right]\,\Theta(r_{c,\mathrm{x}} - r_{ij}),
    \label{eq:ljcut}
\end{equation}
where $r_{ij}$ is the distance between beads $i$ and $j$, $\Theta$ is again the Heaviside step function, $r_{c,\mathrm{x}}$ is the cutoff distance for the respective pair interaction, and
\begin{equation}
    V_{\mathrm{LJ}}(r_{ij}) = 4\varepsilon_{\mathrm{x}}\left[ \left(\frac{\sigma_{\mathrm{x}}}{r_{ij}}\right)^{12} - \left(\frac{\sigma_{\mathrm{x}}}{r_{ij}}\right)^{6} \right],
    \label{eq:lj}
\end{equation}
with $\varepsilon_{\mathrm{x}}$ and $\sigma_{\mathrm{x}}$ the interaction strength and diameter for pair type $\mathrm{x}\in\{\mathrm{dd},\mathrm{dp},\mathrm{pp},\mathrm{dw}\}$ (Table~\ref{tab:ljparams}). The constant shift $-V_{\mathrm{LJ}}(r_{c,\mathrm{x}})$ ensures that the potential vanishes continuously at the cutoff. For purely repulsive interactions we use the Weeks--Chandler--Andersen potential, i.e., $r_{c,\mathrm{x}} = 2^{1/6}\sigma_{\mathrm{x}}$.

\begin{table}[t]
\centering
\caption{Lennard--Jones parameters of the non-bonded pair interactions (reduced units, $\varepsilon_{\mathrm{x}}$ in $\kt$). The DNA--DNA and DNA--wall interactions are purely repulsive; the DNA--protein and protein--protein interactions are attractive. Proteins do not interact with the tether wall.}
\label{tab:ljparams}
\begin{tabular}{lccc}
\toprule
pair (x) & $\varepsilon_{\mathrm{x}}$ $(\kt)$ & $\sigma_{\mathrm{x}}$ & $r_{c,\mathrm{x}}$ \\
\midrule
DNA--DNA (dd)         & $1.0$ & $1.2\,\sigma$ & $2^{1/6}\cdot 1.2\,\sigma$ \\
DNA--protein (dp)     & $\{0.6,0.8\}$ & $2.0\,\sigma$ & $5\,\sigma$ \\
protein--protein (pp)  & $1.5$ & $2.0\,\sigma$ & $5\,\sigma$ \\
DNA--wall (dw)         & $1.0$ & $4.0\,\sigma$ & $2^{1/6}\cdot 4.0\,\sigma$ \\
\bottomrule
\end{tabular}
\end{table}

DNA--DNA as well as DNA--wall interactions are purely repulsive, while protein--protein and protein--DNA interactions feature attraction. The parameter choices for the various pair interactions are summarized in Table~\ref{tab:ljparams}. For DNA--protein interactions, we considered two different values, $0.6\,\kt$ and $0.8\,\kt$. The results shown in the main text were obtained with the latter value; results for the former and for further protein volume fractions are shown in Sec.~\ref{sec:app_conditions}.

\subsection{Tethering geometry and applied tension}
The magnetic tweezers geometry is realized with a fixed bottom anchor and a mobile top wall (Fig.~\ref{fig:App_MD}b,c). The first DNA bead is fixed in both position and orientation and serves as the bottom anchor, and permeation of DNA monomers below its $z$-coordinate is prevented by a repulsive 9--3 LJ potential that spans the transverse width of the box. The last DNA bead is rigidly attached to a honeycomb wall---a rigid monolayer of wall beads arranged on a hexagonal lattice that spans the periodic $x$--$y$ cross-section---with the wall and terminal bead integrated into a single rigid body that is free to translate while its orientation, including the intrinsic orientation of the attached DNA bead, is fixed. This wall prevents linking-number leakage through the chain end; together with the rotational clamping of both termini, the DNA linking number is conserved, and the imposed supercoiling is maintained throughout the run. A constant stretching force $f$ is applied to the rigid body along $+z$. The moving wall interacts with the DNA through a purely repulsive potential, but to avoid piston-compression of the LJ protein gas, it has no physical interaction with the proteins, which can therefore freely permeate it. The box is strongly elongated along $z$, sized to accommodate the full end-to-end extension of the initial, Monte Carlo-generated conformation (see below); it is periodic in the transverse $x$ and $y$ directions and bounded by reflective walls along the stretching axis $z$.

\subsection{Simulation initialization via MC-generated configurations}
To reduce the equilibration time and the necessary $z$-extension of the simulation box, initial DNA configurations are generated with Monte Carlo simulations~\cite{vand22,skor22}, conducted under the same conditions as the MD simulations except for the proteins—same discretization length, elastic stiffnesses ($A$ and $C$), stretching force, and imposed linking number. These configurations are rescaled to LJ units, and a box is generated around them with margin layers in all directions.

The box length along the stretching ($z$) axis is set to the larger of $\langle z \rangle + 4\,s_z$ and $1.25\,z_{\max}$, where $s_z$ and $z_{\max}$ are the standard deviation and maximum of the end-to-end extension sampled in the Monte Carlo run (subject to a lower bound of $0.25\,N\sigma$, with $N$ the number of DNA beads). The two lateral dimensions are fixed to a constant fraction of the contour length, $L_x = L_y \approx 0.16\,N\sigma$, and are enlarged if the configuration protrudes beyond them.

Following box construction, proteins are placed at random, non-overlapping positions at a fixed target volume fraction $\varphi$, the protein number following from the box volume. The protein phase behavior and the role of $\varphi$ in polymer-assisted condensation are discussed separately in Sec.~\ref{sec:lj}.

\subsection{Equations of motion and sampling}
The dynamics of the system is integrated via a Langevin integrator, which numerically solves the Langevin equation
\begin{equation}
    m_i \ddot{\vec{r}}_i = -\nabla_i U - \gamma_i \dot{\vec{r}}_i + \sqrt{2 \kt \gamma_i}\,\bm{\eta}_i,
    \label{eq:langevin}
\end{equation}
where $U$ is the total potential energy, $m_i$ and $\gamma_i$ are the mass and friction coefficient of bead $i$, and $\bm{\eta}_i$ is a Gaussian white noise with $\langle \eta_{i\alpha}(t) \rangle = 0$ and $\langle \eta_{i\alpha}(t)\,\eta_{j\beta}(t') \rangle = \delta_{ij}\,\delta_{\alpha\beta}\,\delta(t-t')$, where $\alpha, \beta \in \{x, y, z\}$ label Cartesian components. The ellipsoidal DNA beads additionally carry rotational degrees of freedom, which are thermostatted through their angular momenta. We integrate with a timestep $\Delta t = 0.01\,\tau_{\mathrm{LJ}}$ and damping time $\tau_{\mathrm{damp}} = 1.0\,\tau_{\mathrm{LJ}}$ at reduced temperature $T = 1$, where $\tau_{\mathrm{LJ}}$ is the LJ time unit. Each system is equilibrated for $1.5 \times 10^{6}\,\tau_{\mathrm{LJ}}$ ($1.5 \times 10^{8}$ steps) and sampled over a production run of $8.5 \times 10^{6}\,\tau_{\mathrm{LJ}}$ ($8.5 \times 10^{8}$ steps), recording bead positions and orientations every $2.5 \times 10^{3}\,\tau_{\mathrm{LJ}}$.

\section{Polymer-assisted condensation and the Lennard--Jones fluid}
\label{sec:lj}
In this work, we consider the phase demixing behavior of a species of proteins with mutually attractive interactions. If mutual attractions are sufficient, there is a range of concentrations for which proteins spontaneously phase separate into a dense liquid phase and a dilute vapor phase. In this case, phase separation occurs irrespective of the DNA and its topology, and condensates can form by spontaneous protein--protein phase separation alone. As we are interested in the mutual interplay between DNA topology and condensation, we study conditions under which the attraction is sufficient for phase separation, but the concentration is not. In this regime, the presence of the DNA can induce condensation via the process of polymer-assisted condensation (PAC) \cite{sommer22pac}.
In this scenario, the attraction of the proteins to the DNA raises the local protein concentration near the DNA into the two-phase region, and a condensate forms that contains the relevant DNA sections. In the current system, these are the sections that do not feel the presence of the external tension, i.e.,~the plectonemes. Remarkably, PAC droplets stop growing once they contain all the corresponding polymer sections.

As proteins are represented by a simple Lennard--Jones (LJ) fluid in our simulations, we are interested in its phase behavior—specifically the critical temperature $T_c^*$, and the location of the liquid--vapor binodal at our operating temperature, i.e., the saturated-vapor concentration (equivalently volume fraction) below which the fluid is a stable single-phase gas. In this section, we show that our simulations were performed under conditions outside the two-phase (vapor--liquid coexistence) region of the truncated-and-shifted Lennard--Jones (LJTS) fluid.

The protein--protein interaction is modeled by the Lennard--Jones potential \Eqref{eq:ljcut}, with $\varepsilon_{\mathrm{pp}} = 1.5$, $\sigma_{\mathrm{pp}} = 2.0$, and $r_{c,\mathrm{pp}} = 2.5\,\sigma_{\mathrm{pp}}$ (all expressed in reduced Lennard--Jones units). At simulation temperature $T = 1.0$, the reduced temperature characterizing the protein--protein interaction is
\begin{equation}
T^* = \frac{\kt}{\varepsilon_{\mathrm{pp}}} = \frac{1.0}{1.5} \approx 0.667,
\end{equation}
well below the critical temperature of the LJTS fluid, $T_c^* = 1.078$~\cite{Vrabec2006}, at which the liquid and vapor branches merge at the critical density $\rho_c^* \approx 0.319$, corresponding to a critical packing fraction $\varphi_c = (\pi/6)\rho_c^* \approx 0.167$. Operating well below $T_c^*$ places the system in a regime where a broad vapor--liquid coexistence region exists, and the binodal has opened up substantially.

At temperature $T^* = 0.667$, interpolating the vapor-branch binodal density of the LJTS fluid reported by Vrabec \emph{et al.}~\cite{Vrabec2006} yields
\begin{equation}
\rho_v^*(T^* = 0.667) = 0.00525 \pm 0.00004.
\end{equation}
Conversion to volume fraction yields
\begin{equation}
\varphi_v = \frac{\pi}{6}\rho_v^* \approx 0.00275,
\label{eq:phi_binodal}
\end{equation}
which is the threshold below which the homogeneous protein gas is the thermodynamically stable single phase.

In our simulations, we consider protein volume fractions $\varphi \leq 0.0025$, i.e., all our simulations lie below the binodal vapor volume fraction $\varphi_v$ (at most $\varphi/\varphi_v \approx 0.9$). The system is therefore in the single-phase gas regime in the absence of DNA, and any condensate observed in the full system is consequently driven by the DNA--protein interaction.
Note that we perform simulations with a fixed number of proteins. This means that upon condensate formation, the protein volume fraction in the bulk moves even further away from the binodal. Thus, condensates are a clear indicator of the PAC mechanism. Nevertheless, we chose the system size to be sufficiently large to avoid significant depletion of the bulk protein concentration upon condensate formation, such that the arrested growth of the condensate is dominated by topology-induced substrate differentiation rather than protein depletion.

\section{Theory}
\label{sec:app_theory}
We model DNA topology-driven co-condensation on a tether held at constant force and fixed linking number by combining the two-phase theory for DNA plectonemes~\cite{mark07,skor22,vand22,sege25} with the phenomenological co-condensate theory introduced by Quail et al.~\cite{Quail2021}. In the sections below, we first summarize the two-phase model for DNA plectonemes and then extend it to incorporate the formation of co-condensates.

\subsection{Plectoneme two-phase model}
\label{sec:twophase}

\begin{figure}[t]
\centering
\includegraphics[width=8.6cm]{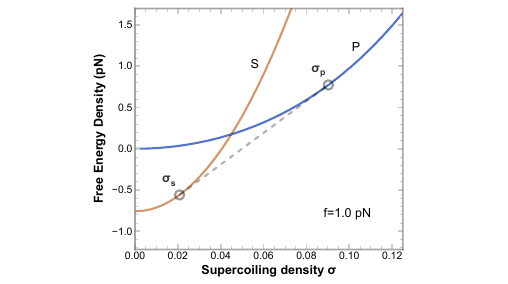}
\caption{
\textbf{Free energy densities of the two-phase model for plectoneme coexistence.}
Free energy densities of the stretched phase, $\sfed$ [\Eqref{eq:def_sfe}; orange], and of the plectonemic phase, $\pfed$ [\Eqref{eq:def_pfe}; blue], at $f = 1.0\,\mathrm{pN}$. The dashed line is the common tangent defined by Eqs.~\eqref{eq:twophase_coex_cond_a} and \eqref{eq:twophase_coex_cond_b}, which touches the two curves at the transition points $\sigs$ and $\sigp$ (circles). For $\sigs < \sig < \sigp$, the free energy density of the molecule follows this tangent.
}

\label{fig:app_twophasefe}
\end{figure}

The two-phase model is a mean-field theory for DNA molecules held under constant tension in the fixed linking number ensemble, which is the relevant ensemble for magnetic tweezer experiments. A single DNA molecule of length $L$, subject to a stretching force $f$ and maintained at constant supercoiling density $\sigma$, is described as partitioned into a stretched, elongated phase and a coiled plectonemic phase, either of which may also span the entire molecule. The elastic behavior of the DNA is described by the twistable worm-like chain (TWLC) model, with bending stiffness $A=42\,\mathrm{nm}$ and torsional stiffness $C=86~\mathrm{nm}$.

In the stretched phase, the majority of the contained linking number is absorbed as twist, as bending fluctuations are attenuated by the force. For sufficiently large forces ($f\gtrsim 0.3\,\mathrm{pN}$ for the typical literature value of the DNA bending persistence length of $50\,\mathrm{nm}$) the phenomenology of this phase is well captured by the free energy density
\begin{equation}
    \label{eq:def_sfe}
    \sfed (\phi,f) = - g(f) + \frac{\ceff(f)}{2} \phi^2 \kt \omn^2,
\end{equation}
where $\phi$ is the supercoiling density in the stretched phase, $\omn \approx 1.76\,\mathrm{rad}/\mathrm{nm}$ is the intrinsic twist of the double helix ($\omn=2\pi / h$, with $h$ the helical pitch),
\begin{equation}
    \label{eq:def_g}
    -g(f) = -f\left(1 - \sqrt{\frac{\kt}{Af}} + g_2 \frac{\kt}{Af} \right) + \ho,
\end{equation}
is the high-force expansion of the WLC extension free energy density~\cite{bust94,mark95}, with $g_2$ a phenomenological constant fitted to the simulations (Sec.~\ref{sec:fitting}) and $\ho$ denoting higher-order terms, and
\begin{equation}
    \label{eq:def_ceff}
    \ceff(f) = C \left( 1 - \frac{C}{4A} \sqrt{\frac{\kt}{Af}} \right) + \ho,
\end{equation}
is the effective torsional stiffness of a TWLC under tension~\cite{moro98}, characterized by reduced effective stiffness due to the absorption of torsional strain in writhe.

There are various expressions for the plectoneme free energy, some of which consider the energetic cost of bending within the coiling geometry, entropy loss due to confinement, and the electrostatic repulsion between the two strands in the plectoneme superhelix~\cite{mark95b,eman13, mark12}, while others resort to simpler phenomenological expressions~\cite {mark07,skor22,vand22,sege25}. For simplicity, we adopt the latter approach using the quartic functional form introduced in Ref.~\cite{skor22}
\begin{equation}
    \label{eq:def_pfe}
    \pfed (\psi) = \left( \frac{P_2}{2}\left(\psi\omn\right)^2 + \frac{P_4}{4}\left(\psi\omn\right)^4 \right) \kt,
\end{equation}
with $\psi$ the supercoiling density in the plectonemic phase and $P_2=13.13\,\mathrm{nm}$ and $P_4=140.78\,\mathrm{nm}^3$ phenomenological parameters, which we obtain by fitting to the MD data.

The two free energy densities are plotted in Fig.~\ref{fig:app_twophasefe} for a representative force of $f=1.0\,\mathrm{pN}$. The WLC-extension term ensures that the free energy of the stretched phase is lower than that of the plectonemic phase at low supercoiling densities, such that for sufficiently low $\sig$ the system is fully in the stretched phase. However, at a force-dependent buckling point $\sigs$ it becomes energetically beneficial to absorb a fraction of the chain length into the plectonemic phase; the system undergoes a buckling transition and enters a regime of coexisting stretched and plectonemic domains. Ignoring interfacial terms~\cite{sege25}, the total free energy density is expressed as a length-weighted combination of the two contributions
\begin{align}
    \label{eq:def_ffe}
    \ffed(\sig,f) &= \left(1-\chi\right) \sfed(\phi) + \chi \pfed(\psi),
\end{align}
where $0 \leq \chi \leq 1$ is the fraction of chain length in the plectonemic phase, i.e., for given $\chi$, a length $L_s=(1-\chi)L$ is in the stretched phase and a length $\lplec=\chi L$ is in the plectonemic phase. As we are considering the fixed linking number ensemble, the total linking number contained in the individual phases must sum to the total linking number of the molecule, which is expressed via the constraint
\begin{equation}
    \label{eq:sig_partition}
    \sigma = \left(1-\chi\right) \phi + \chi \psi.
\end{equation}
This constraint implies that there are only two independent order parameters among $\chi$, $\phi$, and $\psi$. The total free energy of a molecule is $\ffe = L\,\ffed$.

Minimizing the free energy \Eqref{eq:def_ffe} shows that the system exhibits two transition points $\sigs < \sigp$. Pure stretched and plectonemic phases spanning the entire molecule are found for $\sig < \sigs$ and $\sig > \sigp$, respectively, while for $\sigs < \sig < \sigp$ the system is in a coexistence regime of stretched and plectonemic domains. The transition points are force-dependent, with both $\sigs$ and $\sigp$ increasing with increasing force. These transition points are defined via the minimization criteria
\begin{align}
    \label{eq:twophase_coex_cond_a}
    \left.\frac{\partial \pfed}{\partial \psi} \right|_{\psi = \sigp} &= \left.\frac{\partial \sfed}{\partial \phi} \right|_{\phi = \sigs},
    \\
    \label{eq:twophase_coex_cond_b}
    \pfed(\sigp) - \sfed(\sigs) &= \left(\sigp - \sigs\right) \left.\frac{\partial \sfed}{\partial \phi} \right|_{\phi = \sigs}.
\end{align}
Within the coexistence regime, the free energy is given by a common tangent construction (see Fig.~\ref{fig:app_twophasefe}), characterized by the condition that the slopes of the two free energy curves are equal at their respective transition points. This translates to equal torque in the two phases at coexistence, as the torque is given by~\cite{mark07,sege25}
\begin{equation}
    \label{eq:twophase_torque}
    \tau = \frac{1}{\omn} \frac{\partial \ffed}{\partial \sig}.
\end{equation}

Similarly, the molecular extension along the direction of the applied force is given by~\cite{mark07,skor22}
\begin{equation}
    \label{eq:twophase_ext}
    z = -L\frac{\partial \ffed}{\partial f}.
\end{equation}

\subsection{Co-condensate model}
\label{sec:cocond}
To describe the formation of co-condensates between semiflexible polymers and phase-separating proteins, we employ the phenomenological model introduced by Quail et al.~\cite{Quail2021},
\begin{equation}
    \label{eq:juelicher_cocond}
    \cfe(L) = - \nu\alpha L + 4 \pi \gamma \left(\frac{3\alpha}{4\pi}\right)^{2/3} L^{2/3},
\end{equation}
where $\nu$ is the condensation free energy per unit volume, $\alpha$ is the condensate volume per unit DNA contour length, and $\gamma$ is the surface tension of the co-condensate. The first term corresponds to the bulk free energy gain from condensation, while the second term corresponds to the interfacial tension. The interfacial term assumes a spherical condensate geometry: anisotropic shapes (e.g., prolate or cylindrical) can be favored when the polymer rigidity competes with the surface tension, but as long as the aspect ratio does not depend on droplet size, such geometric corrections only renormalize the prefactor of the $L^{2/3}$ term and are absorbed into the phenomenological coefficient $b$ introduced below.

To account for the bending rigidity of the polymer and the associated bending energy required to confine the polymer within a spherical droplet, we follow Refs.~\cite{Odijk1983,GrosbergKhokhlov1994,Sakaue2007} and introduce a bending energy contribution of the form
\begin{equation}
    \label{eq:fbend}
    \fbend(L) \sim \kt \frac{AL}{R^2} \sim \kt \frac{AL}{\alpha^{2/3}L^{2/3}} \sim c L^{1/3}.
\end{equation}

Combining the contributions of Eqs.~\eqref{eq:juelicher_cocond} and \eqref{eq:fbend} we arrive at the following phenomenological form for the co-condensate free energy
\begin{equation}
    \label{eq:def_cfewlc}
    \cfe(L) = -a L + b L^{2/3} + c L^{1/3},
\end{equation}
where the constants $a,b,c \geq 0$ consolidate the multiplicative constants from the above expressions.

Formation of such a co-condensate is favorable if $\cfe(L) < 0$. For small $L$, the interfacial and bending contributions dominate, prohibiting co-condensation, while for large $L$, the bulk gain dominates. The critical length beyond which co-condensation becomes favorable is given by
\begin{equation}
    \label{eq:def_Lc}
    L_c = \left(\frac{b + \sqrt{b^2 + 4ac}}{2a}\right)^3.
\end{equation}
In the simpler case where $c=0$ this reduces to $L_c = (b/a)^3$.

\subsection{Two-phase model with co-condensation}
\label{sec:pleccond}
To combine the phenomenology of co-condensation with that of DNA under tension and topological constraints, we make the following assumptions. First, we assume all plectonemic phase length $\lplec$ to be contained in a single droplet. This is supported by our simulations, in which co-condensed plectonemes coarsen into a single droplet [Fig.~\ref{fig:fig4}], and follows from the sub-additivity of the interfacial and bending terms in \Eqref{eq:def_cfewlc}: since $(L_1+L_2)^{2/3} < L_1^{2/3}+L_2^{2/3}$, and likewise for the exponent $1/3$, splitting a droplet at fixed DNA content always raises its free energy. Second, the diameter of the co-condensate droplet is assumed to be small compared to the contour length of the DNA contained in that droplet, such that the entry--exit point distance is negligible and the DNA contained in the droplet is effectively insulated from the stretching force. This then suggests that the part of the co-condensate free energy corresponding to the DNA contained in the droplet is well described by the plectoneme free energy density $\pfed(\psi)$. This assumption holds well for large supercoiling densities where the DNA is plectonemically coiled and its entry and exit points nearly coincide.

We therefore write the free energy of a co-condensate containing DNA of length $\lplec$ as
\begin{equation}
    \label{eq:def_cfe}
    F_{\mathrm{cond}}(\psi,\lplec) = \lplec\,\pfed(\psi) +\cfe(\lplec)\,\Theta\left(-\cfe(\lplec)\right),
\end{equation}
where $\Theta$ is again the Heaviside step function, which enforces the inclusion of the co-condensate term only if it is energetically favorable, i.e., when $\cfe(\lplec) < 0$.

To describe the full system consisting of stretched phase and potentially co-condensed plectonemic phase, we write the total free energy in the same two-phase decomposition as before ($L_s = (1-\chi)L$ and $\lplec = \chi L$), but now with the free energy of the plectonemic phase modified by the co-condensate contribution, i.e.,
\begin{align}
    \label{eq:def_cond_ffe}
    \ffe(\sig,f,L) = &\phantom{+} (1-\chi) L \sfed(\phi,f)
    \n
    &+ \chi L \pfed(\psi) + \cfe(\chi L) \,\Theta\left(-\cfe(\chi L)\right).
\end{align}

As in the condensate-free model, torque and extension follow from Eqs.~\eqref{eq:twophase_torque} and \eqref{eq:twophase_ext} with $\ffed = \ffe/L$. Depending on $\sig$ and $f$, the system exhibits up to four regimes: a pure stretched phase; a coexistence regime of stretched and bare plectonemic domains, which persists from the pure two-phase model wherever co-condensation is not yet favorable; a coexistence regime of stretched and co-condensed plectonemic domains; and a pure co-condensed plectonemic phase. Within the co-condensed coexistence regime, the free energy is again given by a linear interpolation between the stretched phase free energy on the one side and the combined plectoneme and co-condensate free energy on the other. However, in the pure two-phase model the supercoiling densities at which this interpolation terminates, $\sigs$ and $\sigp$, are fixed once by the common tangent construction and remain unchanged throughout the coexistence regime, such that any variation in linking number manifests purely in conversion of contour length between the two phases at constant supercoiling densities within each phase. The sub-extensive contributions of the condensate free energy invalidate this simple construction and lead to $\chi$-dependent variations of the two supercoiling densities $\sigs^*(\chi)$ and $\sigp^*(\chi)$ in the two phases.

Minimization of the free energy \Eqref{eq:def_cond_ffe} for given $\sig$ and $f$, subject to the partitioning constraint \Eqref{eq:sig_partition}, leads to the minimization conditions
\begin{align}
    \label{eq:cond_coex_cond_a}
    \left.\frac{\partial \pfed}{\partial \psi} \right|_{\psi = \sigp^*} &= \left.\frac{\partial \sfed}{\partial \phi} \right|_{\phi = \sigs^*},
    \\
    \label{eq:cond_coex_cond_b}
    \pfed(\sigp^*) - \sfed(\sigs^*) &= \left(\sigp^* - \sigs^*\right) \left.\frac{\partial \sfed}{\partial \phi} \right|_{\phi = \sigs^*}
    \n
    &\phantom{=} - \frac{1}{L}\left.\frac{\partial \cfe}{\partial \chi} \right|_{\chi = \chi^*} \Theta\left(-\cfe(\chi^* L)\right),
\end{align}
where the Heaviside step function restricts the last term to droplet sizes for which co-condensation is favorable, i.e., to $\chi^* L > L_c$ [\Eqref{eq:def_Lc}]. This term is the marginal change in free energy upon growing the droplet by a unit of contour length, and explicitly reads
\begin{equation}
    \label{eq:cfe_marginal}
    \frac{1}{L}\left.\frac{\partial \cfe}{\partial \chi}\right|_{\chi = \chi^*}
    = -a + \frac{2b}{3\left(\chi^* L\right)^{1/3}} + \frac{c}{3\left(\chi^* L\right)^{2/3}}.
\end{equation}

Of the three contributions, only the bulk term $-a$ is independent of $\chi$. By itself, it would amount to a constant shift of the plectonemic free energy density, which merely shifts the two transition points $\sigs$ and $\sigp$, leaving the ordinary common tangent construction in play. The remaining—interfacial and bending—contributions, however, depend on $\chi$ and therefore elicit a $\chi$-dependent rebalancing of the coexistence conditions. The supercoiling densities of the two phases consequently assume distinct $\chi$-dependent values $\sigs^*(\chi)$ and $\sigp^*(\chi)$.

For given $\sig$, the partitioning is then determined self-consistently, with $\chi^*$ simultaneously satisfying both coexistence conditions and the constraint $\sig = \left(1-\chi^*\right)\sigs^*(\chi^*) + \chi^*\sigp^*(\chi^*)$. The coexistence regime is consequently no longer described by a common tangent construction. Instead, as $\sig$ increases and the droplet grows, the marginal free energy \Eqref{eq:cfe_marginal} converges toward its bulk value $-a$, and both $\sigs^*$ and $\sigp^*$ shift accordingly.
Thus, the free energy remains an interpolation between the two phases, but between endpoints that themselves move as the coexistence regime is traversed. The free energy is no longer linear in $\sig$ and the torque is no longer constant, but rebalances continuously. As $\chi$ appears in the denominators of \Eqref{eq:cfe_marginal}, the effect of the sub-extensive contributions is most pronounced for small droplets, i.e., just beyond the buckling transition, and becomes negligible for very long tethers and large $\sigma$ [the contribution decays as $(\chi L)^{-1/3}$], where the ordinary construction with uniformly displaced transition points is recovered.
Moreover, the step function in \Eqref{eq:cond_coex_cond_b} implies that the marginal free energy \Eqref{eq:cfe_marginal} enters discontinuously: it jumps from zero to a finite negative value ($-a/3$ for $c = 0$) once the droplet reaches the minimum stable size $L_c$, such that the coexistence conditions cannot be satisfied continuously across this point. Condensates instead appear at finite size, giving rise to jumps in $\chi$ and corresponding discontinuities in the torque and extension curves, which are signatures of the co-condensation process.

Depending on the balance between tension, linking number, and condensate properties, the formation of the condensate may occur for supercoiling densities smaller than the condensate-free $\sigs$ or above it. In the former case, the condensate formation shifts the observed buckling transition to lower $\sigma$. Conversely, in the latter case, the buckling transition is unaffected by the proteins and initially forms bare plectonemes, which may then undergo a secondary transition into a co-condensate at some $\sig > \sigs$.

\subsection{Solving the two-phase models}
\label{sec:numsolve}
The general strategy for finding the two-phase free energy of either model is to first solve the corresponding coexistence conditions for the transition points ($\sigs$ and $\sigp$, or $\sigs^*$ and $\sigp^*$), and then use \Eqref{eq:sig_partition} to substitute the result into the free energy, \Eqref{eq:def_ffe} or \Eqref{eq:def_cond_ffe}, respectively. All observables are then obtained from the free energy by differentiation; see for example Eqs.~\eqref{eq:twophase_torque} and \eqref{eq:twophase_ext}.

For the free energy densities $\sfed$ and $\pfed$ as given in Eqs.~\eqref{eq:def_sfe} and \eqref{eq:def_pfe}, respectively, the coexistence conditions of the condensate-free two-phase model \Eqref{eq:twophase_coex_cond_a} and \Eqref{eq:twophase_coex_cond_b} permit a closed form solution, albeit an unwieldy one due to the cubic roots involved. However, inclusion of the co-condensate term renders the corresponding conditions \Eqref{eq:cond_coex_cond_a} and \Eqref{eq:cond_coex_cond_b} inaccessible to an analytical treatment altogether. We therefore solve both models numerically, following the approach outlined below. It presupposes no particular functional form and requires only that the two free energy densities be convex in their respective supercoiling densities and minimal at vanishing supercoiling, as is the case for Eqs.~\eqref{eq:def_sfe} and \eqref{eq:def_pfe}.

\subsubsection{Two-phase model}
\label{sec:numsolve_bare}
For the ordinary two-phase model, the transition points are independent of $\sig$, and thus need to be determined only once per force.
The two conditions are solved by nesting two one-dimensional problems. For a trial plectonemic density $\psi$, the equal torque condition \Eqref{eq:twophase_coex_cond_a} fixes the corresponding stretched phase density $\phi(\psi)$, which is unique by the convexity of $\sfed$ and bounded from above by $\psi$, as the plectonemic phase is always the more strongly supercoiled one. The tangency condition \Eqref{eq:twophase_coex_cond_b} thereby reduces to a scalar equation in the single remaining variable $\psi$, with residual
\begin{equation}
    \label{eq:num_resid_bare}
    r(\psi) = \sfed\left(\phi(\psi)\right) - \pfed(\psi) + \left(\psi - \phi(\psi)\right) \left.\frac{\partial \sfed}{\partial \phi}\right|_{\phi = \phi(\psi)},
\end{equation}
whose root yields the two transition points as $\sigp = \psi$ and $\sigs = \phi(\sigp)$. As long as $\pfed$ is convex, this root can be shown to be unique on the admissible domain $\phi \leq \psi$, irrespective of the particular functional forms chosen; the remaining roots of $r(\psi)$ correspond to $\phi > \psi$ and are excluded by construction. All involved derivatives are evaluated by complex-step differentiation to avoid finite-difference errors.

For given $\sig$, the fraction of the molecule contained in the plectonemic phase follows from the partitioning constraint \Eqref{eq:sig_partition} as
\begin{equation}
    \label{eq:num_chi_bare}
    \chi = (\sig - \sigs)/(\sigp - \sigs),
\end{equation}
which assumes values between 0 and 1 within the coexistence regime. Insertion of \Eqref{eq:num_chi_bare}, $\sigs$, and $\sigp$ into the free energy \Eqref{eq:def_ffe} yields the two-phase free energy. Observables are then computed following Eqs.~\eqref{eq:twophase_torque} and \eqref{eq:twophase_ext}.

\subsubsection{Two-phase model with co-condensation}
\label{sec:numsolve_cond}
The condensate term in \Eqref{eq:def_cond_ffe} renders the common tangent construction $\sig$-dependent, as the last term of \Eqref{eq:cond_coex_cond_b} contains $\chi^*$, which is itself fixed by the partitioning constraint at the imposed $\sig$. The conditions therefore have to be solved for every individual $\sig$, and $\sigs^*$ and $\sigp^*$ become functions of the imposed supercoiling density.

The step function splits the free energy into a condensate-free branch, on which the droplet term is absent, and a condensed branch, on which it is active. On the condensed branch the equal torque condition \Eqref{eq:cond_coex_cond_a} is unaffected by the condensate, such that $\phi(\psi)$ is obtained as in Sec.~\ref{sec:numsolve_bare} and the second condition again reduces to a scalar equation in $\psi$, now carrying the marginal condensate term \Eqref{eq:cfe_marginal} evaluated at the $\chi$ fixed by the partitioning constraint at the imposed $\sig$. Through this $\chi$ the additional term depends on $\psi$, so that convexity of the free energy densities no longer constrains the number of solutions; together with the discontinuity of the step function at the minimum droplet size, the residual may possess several roots or none. We therefore identify all roots via bracketing and subsequent refinement.

Stationarity holds within each branch separately and therefore cannot decide between them. Instead, we evaluate the total free energy \Eqref{eq:def_cond_ffe} at the imposed $\sig$ for the condensate-free state, for the partially condensed state belonging to each root, and for the fully condensed state, and select the global minimum. Transitions between branches are marked by discontinuities in torque and extension. These are first-order transitions between distinct droplet states.

\section{Parameter fitting and phase diagram}
\label{sec:fitting}
The two-phase model and the co-condensate extension both depend on explicitly set simulation parameters and on several phenomenological parameters which are fitted to corresponding MD simulations.

Explicitly set parameters are the elastic constants $A=42\,\mathrm{nm}$ and $C=86\,\mathrm{nm}$, the thermal energy $\kt=4.114\,\mathrm{pN}\,\mathrm{nm}$, the intrinsic twist $\omn=1.76\,\mathrm{rad}/\mathrm{nm}$, and the contour length $L=2720\,\mathrm{nm}$ (corresponding to the considered $8$\,kbp molecules).

\subsection{Parameter fitting}
The remaining parameters of the bare two-phase model, $g_2$, $P_2$ and $P_4$, are determined from the rotation curves simulated in the absence of proteins. Both extension and twist density $\sigtw = \dtw / \lkn$ (related to torque via $\tau = C \kt \omn \sig_{\mathrm{tw}}$) are fitted simultaneously for all considered forces ($f=0.5,1,2,4\,\mathrm{pN}$), and all corresponding supercoiling densities. Both residuals are normalized by the root-mean-square of the respective data and by the square root of the number of data points, and both enter the cost with equal weight, as do the four forces. Least-squares minimization yielded the values $g_2 = 0.72$, $P_2 = 13.13\,\mathrm{nm}$ and $P_4 = 140.78\,\mathrm{nm}^3$.

The condensate parameters $a$, $b$ and $c$ of \Eqref{eq:def_cfewlc} are obtained in the same manner from the rotation curves simulated in the presence of proteins, with all remaining parameters treated as constants and adopted as listed above. Since the condensate free energy is piecewise flat in $b$ and $c$ (these parameters enter the model only through the selected coexistence construction), pure gradient-based optimization is insufficient. Minimization therefore proceeds in two stages: a grid scan of $(a,b,c)$ followed by a local derivative-free (Nelder--Mead) refinement. Every combination of protein volume fraction $\varphi$ and DNA--protein interaction strength $\varepsilon_{\mathrm{dp}}$ is fitted separately.

In practice we observe the parameter space to be degenerate, in particular in $b$ and $c$. In all cases the inclusion of only $b$ proves sufficient to capture the observed behavior, and we therefore set $c=0$ for all fits. For the three largest considered forces, a joint fit of $a$ and $b$ yields good agreement with the MD data, while the lowest force $f=0.5\,\mathrm{pN}$ requires separate parameters, in particular to account for the sharp extension collapse that occurs before buckling, at small but finite $\sig$. For the simulations shown in the main text ($\varphi = 0.0010$ and $\varepsilon_{\mathrm{dp}} = 0.8\,\kt$) the fit yielded $a = 0.78\,\mathrm{pN}$ and $b = 4.06\,\mathrm{pN}\,\mathrm{nm}^{1/3}$ for the three larger forces and $a = 0.49\,\mathrm{pN}$ and $b = 2.98\,\mathrm{pN}\,\mathrm{nm}^{1/3}$ for $f=0.5\,\mathrm{pN}$.

All model parameters are summarized in Table~\ref{tab:fitparams}.
\begin{table}[t]
\centering
\caption{Parameters of the two-phase model and of its co-condensate extension. The simulation inputs are set to the values used in the MD simulations, while all remaining parameters are obtained by the fitting procedure outlined in Section~\ref{sec:fitting}. The bare parameters $g_2$, $P_2$ and $P_4$ are fitted jointly to all forces of the protein-free data set. The condensate parameters $a$, $b$ and $c$ refer to $\varphi = 0.0010$ and $\varepsilon_{\mathrm{dp}} = 0.8\,\kt$, and are fitted jointly for $f\geq 1\,\mathrm{pN}$ and separately for $f = 0.5\,\mathrm{pN}$. As $b$ and $c$ are degenerate, $c$ is set to zero throughout.}
\label{tab:fitparams}
\begin{tabular}{lcl}
\toprule
parameter & value & description \\
\midrule
\multicolumn{3}{l}{{simulation inputs}} \\
$A$    & $42\,\mathrm{nm}$                      & bending stiffness \\
$C$    & $86\,\mathrm{nm}$                      & torsional stiffness \\
$\kt$  & $4.114\,\mathrm{pN}\,\mathrm{nm}$      & thermal energy \\
$\omn$ & $1.76\,\mathrm{rad}/\mathrm{nm}$       & intrinsic twist \\
$L$    & $2720\,\mathrm{nm}$                    & contour length ($8\,\mathrm{kbp}$) \\
\midrule
\multicolumn{3}{l}{{two-phase model, all $f$}} \\
$g_2$ & $0.72$              & WLC extension, \Eqref{eq:def_g} \\
$P_2$ & $13.13\,\mathrm{nm}$  & plectoneme free energy, \Eqref{eq:def_pfe} \\
$P_4$ & $140.78\,\mathrm{nm}^3$ & plectoneme free energy, \Eqref{eq:def_pfe} \\
\midrule
\multicolumn{3}{l}{{co-condensate, $f \geq 1\,\mathrm{pN}$}} \\
$a$ & $0.78\,\mathrm{pN}$                       & bulk gain, \Eqref{eq:def_cfewlc} \\
$b$ & $4.06\,\mathrm{pN}\,\mathrm{nm}^{1/3}$    & interfacial penalty, \Eqref{eq:def_cfewlc} \\
$c$ & $0$                                       & bending penalty, \Eqref{eq:def_cfewlc} \\
\midrule
\multicolumn{3}{l}{{co-condensate, $f = 0.5\,\mathrm{pN}$}} \\
$a$ & $0.49\,\mathrm{pN}$                       & bulk gain, \Eqref{eq:def_cfewlc} \\
$b$ & $2.98\,\mathrm{pN}\,\mathrm{nm}^{1/3}$    & interfacial penalty, \Eqref{eq:def_cfewlc} \\
$c$ & $0$                                       & bending penalty, \Eqref{eq:def_cfewlc} \\
\bottomrule
\end{tabular}
\end{table}

\subsection{Phase diagram construction}
The phase diagram in Fig.~\ref{fig:fig2}(d) is constructed from the model using the parameters described in the previous subsection. As the condensate parameters $a$ and $b$ exhibit force dependence for the lowest considered force, the obtained values are interpolated linearly between, and extrapolated beyond, the fitted forces. Extrapolation below $0.5\,\mathrm{pN}$ is performed by continuing the line through the two lowest fitted forces. The transition points $\sigs$ and $\sigp$, marking the buckling transition and the transition into the fully plectonemic phase, respectively, are evaluated as outlined in Sec.~\ref{sec:numsolve_bare}. For a given force, the onset of co-condensate formation [green line in Fig.~\ref{fig:fig2}(d)] is determined as the smallest supercoiling density $\sig$ for which the co-condensate free energy \Eqref{eq:def_cond_ffe} is smaller than the condensate-free free energy $L\,\ffed$ [\Eqref{eq:def_ffe}]. Transition into full absorption of the molecule into the co-condensate [orange line in Fig.~\ref{fig:fig2}(d)] is then defined as the smallest $\sig$ for which the co-condensate fraction $\chi^*$ reaches unity. Both boundaries are evaluated on a grid of forces and supercoiling densities, and the resulting points are smoothed to obtain continuous curves.

\section{Rotation curves at further protein conditions}
\label{sec:app_conditions}
\begin{figure*}[t]
\centering
\includegraphics[width=17.8cm]{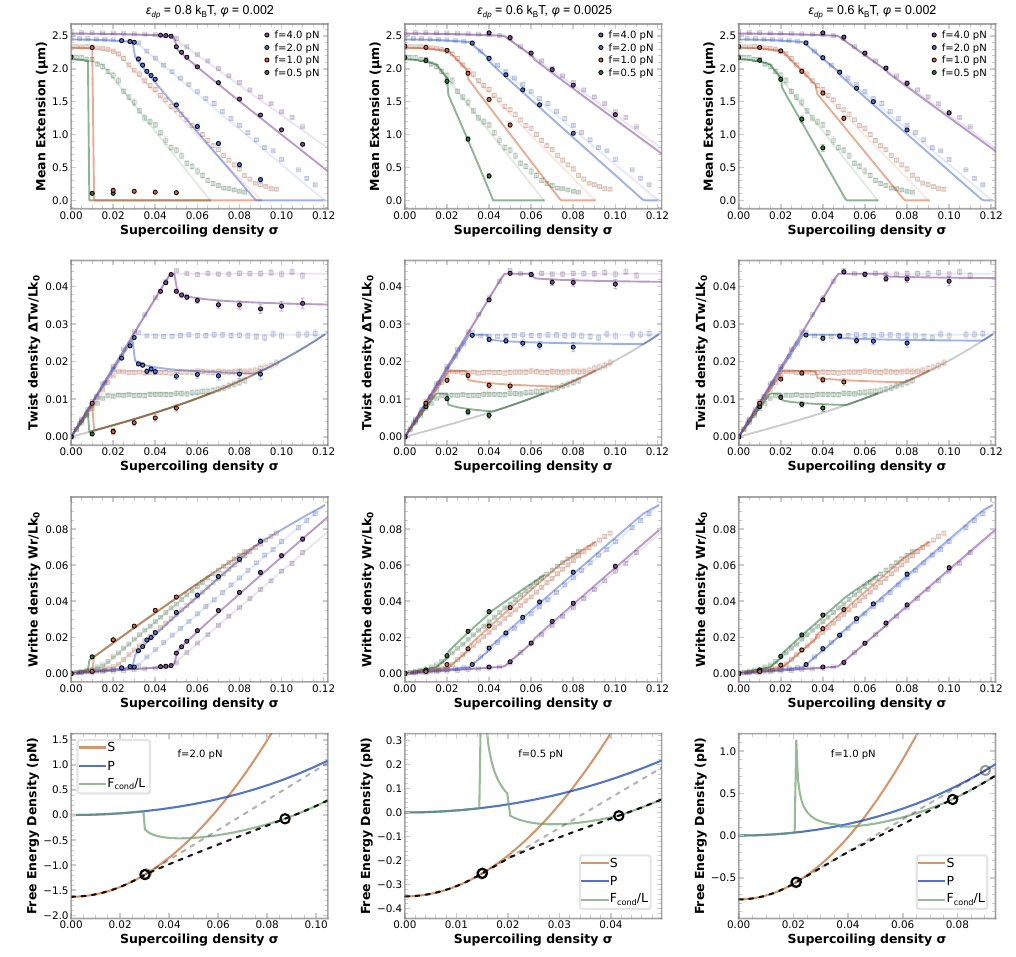}
\caption{
\textbf{Rotation curves at further protein conditions.}
Columns: $\varepsilon_{\mathrm{dp}} = 0.8\,\kt$, $\varphi = 0.002$ (left); $\varepsilon_{\mathrm{dp}} = 0.6\,\kt$, $\varphi = 0.0025$ (middle); $\varepsilon_{\mathrm{dp}} = 0.6\,\kt$, $\varphi = 0.002$ (right). Rows 1--3: mean extension, twist density $\dtw/\lkn$, and writhe density $\wr/\lkn$ versus supercoiling density $\sig$ at $f = 0.5$, $1.0$, $2.0$, and $4.0\,\mathrm{pN}$. Squares: protein-free simulations; circles: with proteins; transparent and saturated lines: two-phase model without and with co-condensation, respectively. Row 4: free energy densities $\sfed$, $\pfed$, and $\ffe_{\mathrm{cond}}/L$ at the indicated force, with the protein-free common tangent (gray dashed), the minimized free energy density of the co-condensate model (black dashed), and the coexistence boundaries (circles).
}
\label{fig:app_rotcurves}
\end{figure*}

Rotation curve simulations conducted with different combinations of DNA--protein attraction and protein volume fraction are shown in Fig.~\ref{fig:app_rotcurves}. In particular, at the weaker DNA--protein attraction the sub-extensive interfacial term delays condensation beyond the buckling point, so that bare plectonemes form first and co-condense only at larger $\sig$.

\section{Plectoneme and droplet analysis}
\label{sec:app_dropfinder}
Plectonemic regions and protein droplets are identified in individual MD snapshots. The former analysis is performed using the method previously applied in Refs.~\cite{skor22,sege25}, which identifies plectonemic regions based on the writhe map (the pairwise contributions of DNA segments to the writhe of the configuration) of individual configurations.

\subsection{Droplet identification and size measurement}
To automate detection of condensates and to facilitate quantitative analysis of the number of droplets and their size in individual simulation snapshots, we employ a custom Python analysis tool \footnote{\url{https://github.com/eskoruppa/DropFinder}} that detects condensates based on the local protein density. The procedure consists of three steps: (i) construction of a smooth protein density field on a regular mesh, (ii) calibration of a density threshold that separates the dilute gas from the condensed phase, and (iii) identification of connected condensate regions.

\subsubsection{Rasterized density field}
For a snapshot with protein positions $\vec{r}_i$, the simulation box, of side lengths $L_\alpha$ along the axes $\alpha \in \{x, y, z\}$, is discretized into a regular mesh of cells of edge lengths $\Delta_\alpha = L_\alpha / N_\alpha$ with $N_\alpha = \lceil L_\alpha / \Delta \rceil$, i.e., the nominal mesh spacing $\Delta = 1\,\sigma$ is adjusted per axis so that an integer number of cells fills the box. Each protein is distributed over the mesh with a normalized, truncated Gaussian kernel,
\begin{equation}
    w(\vec{r}) \propto \exp\!\left(-\frac{\|\vec{r}\|^{2}}{2 s^{2}}\right)\,
    \Theta\!\left(r_{\mathrm{sm}} - \|\vec{r}\|\right),
    \qquad
    s = n_{s}\,\sigma_{\mathrm{prot}},
    \label{eq:dropfinder_kernel}
\end{equation}
of width $s = 3\,\sigma$ (i.e., $n_{s} = 1.5$ times the protein diameter $\sigma_{\mathrm{prot}} = 2\,\sigma$) and truncated at $r_{\mathrm{sm}} = 5\,\sigma$. The resulting field
\begin{equation}
    \rho(\vec{c}) = \sum_{i=1}^{N_{\mathrm{prot}}} w(\vec{r}_{\vec{c}} - \vec{r}_i)
    \label{eq:dropfinder_mesh}
\end{equation}
therefore measures the local protein number per mesh cell, i.e., the local number density in units of the cell volume $\prod_\alpha \Delta_\alpha$. Here, $\vec{c} = (c_x,c_y,c_z)$ with $c_\alpha \in \{0,\dots,N_\alpha-1\}$ is an integer-valued mesh cell label for the cell centered at $\vec{r}_{\vec{c}}$ with components $(c_\alpha + 1/2)\,\Delta_\alpha$. Conversely, $\vec{c}(\vec{r})$ denotes the cell containing the point $\vec{r}$, i.e., $c_\alpha = \lfloor r_\alpha / \Delta_\alpha \rfloor$. Kernel contributions wrap around the periodic boundaries of the simulation box ($x$ and $y$ in the present case; see Sec.~\ref{sec:app_md}) and are discarded where they extend beyond the box along the non-periodic stretching direction ($z$).

\subsubsection{Density threshold and droplet identification}
Individual cells are classified as belonging to the dilute or condensed phase based on their local density $\rho(\vec{c})$. Rather than imposing a fixed threshold, the threshold $\rho_{\mathrm{th}}$ is calibrated for each trajectory based on the observed distribution of local densities. This ensures that the detection criterion adapts to the protein volume fraction $\varphi$ and to the interaction strengths, and requires no manual tuning across the parameter scans.
Every protein is assigned the density of the cell it occupies, $\rho_i = \rho(\vec{c}(\vec{r}_i))$, and the distribution $P(\rho)$ of these per-protein local densities is obtained from all available snapshots (for large trajectories it is usually sufficient to consider subsets of evenly spaced snapshots). In a phase-separated system this distribution is expected to be bimodal: a sharp low-density peak from the surrounding gas and a broad high-density shoulder from proteins inside the condensate.

The threshold is placed in the tail of the gas peak, which requires locating both its position and its width. To this end, $P(\rho)$ is histogrammed with $50$ proteins per bin on average, and $\rho_{\mathrm{gas}}$ is identified as the dominant low-density maximum. Scanning upwards from there, the gas peak is taken to end at the first bin holding less than $2\%$ of the cumulative weight up to that bin or less than $20\%$ of the preceding bin, refined to the nearest local minimum $\rho_{\mathrm{min}}$. The standard deviation $s_{\mathrm{gas}}$ of the distribution truncated at $\rho_{\mathrm{min}}$ then measures the density fluctuations within the gas, and the threshold is set to
\begin{equation}
    \rho_{\mathrm{th}} = \max\!\Big[\rho_{\mathrm{min}},\;
    \min\!\big(\rho_{\mathrm{gas}} + n_{\mathrm{std}}\, s_{\mathrm{gas}},\;
    \rho_{\mathrm{max}}\big)\Big],
    \qquad n_{\mathrm{std}} = 4,
    \label{eq:dropfinder_cutoff}
\end{equation}
with $\rho_{\mathrm{max}}$ the largest sampled density. A cell is thus considered condensed if its density exceeds the typical gas density by four standard deviations of the gas fluctuations, the outer $\max$ and inner $\min$ merely ensuring that the threshold neither falls back into the gas peak nor exceeds the sampled range.

All cells with $\rho(\vec{c}) > \rho_{\mathrm{th}}$ are marked, and connected sets of marked cells are identified as individual droplets. Adjacency includes face-, edge-, and corner-sharing cells and wraps around the periodic directions.

\subsubsection{Condensate size}
The volume $V_{\mathrm{cond}}$ of a condensate follows from its cell count $n_{\mathrm{cond}}$,
\begin{equation}
    V_{\mathrm{cond}} = n_{\mathrm{cond}} \prod_{\alpha} \Delta_\alpha,
    \label{eq:dropfinder_volume}
\end{equation}
i.e., from the volume enclosed by the threshold iso-density surface rather than from the protein positions, so that it is insensitive to the density profile inside the condensate. The condensate size is defined as the radius of the sphere of equal volume,
\begin{equation}
    R_{\mathrm{cond}} = \left(\frac{3 V_{\mathrm{cond}}}{4\pi}\right)^{1/3},
    \label{eq:dropfinder_reff}
\end{equation}
evaluated for the largest droplet of the snapshot, i.e., the one with the largest protein content. The number of droplets per snapshot is obtained by counting all droplets that hold at least $10\%$ of the protein content of the largest one, which discards transient subcritical clusters that continuously form and dissolve in the surrounding gas. Snapshots in which the largest droplet holds fewer than $25$ proteins are classified as droplet-free (zero droplets).

\section{Magnetic tweezers experiments}
\label{sec:app_exp}

\begin{figure*}[t]
\centering
\includegraphics[width=12.9cm]{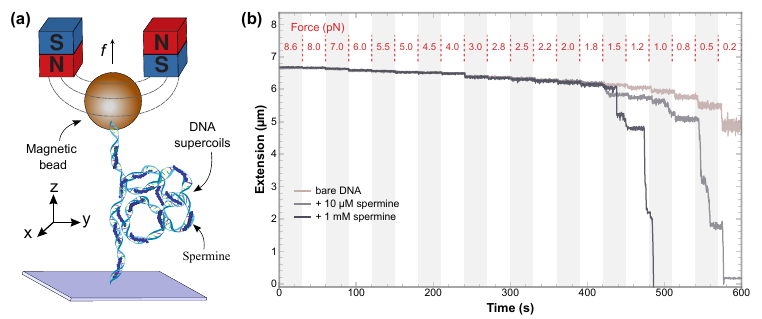}
\caption{
\textbf{Magnetic tweezers measurements of spermine-induced DNA collapse.}
(a) Schematic of the magnetic tweezers measurement. A DNA molecule is tethered between a flow cell surface and a superparamagnetic bead. External magnets exert a stretching force $f$ and control the rotation of the bead and, therefore, the linking number of the DNA. Spermine interacts with DNA and leads to collapse of the tether.
(b) Extension-time traces of 20.6 kbp DNA tethers at different stretching forces (indicated on top) in the presence of 0, 10~$\mu$M and 1~mM spermine (light to dark traces). As the force is lowered, the DNA extension shortens slightly, following the well-known WLC response. In the presence of spermine, the DNA tethers initially follow the same response as in the absence of spermine at high forces. However, as the force is lowered, rapid compaction steps are observed, eventually leading to complete collapse of the DNA tether (at $\sim$1.2 pN for the 1 mM spermine condition and at $\sim$0.5 pN for the 10 $\mu$M condition).
}
\label{fig:SI_exp_timetraces}
\end{figure*}

\subsection{Materials}

For the measurements, we used a $20.6$\,kbp dsDNA construct labeled at one end with multiple digoxigenin and at the other with multiple biotin groups~\cite{lipf09b,lipf10b}. DNA--bead attachment was performed by incubating $2~\mu\mathrm{L}$ of picomolar DNA stock with $2~\mu\mathrm{L}$ of streptavidin-coated superparamagnetic beads (1.05~$\mu\mathrm{m}$ diameter MyOne beads, Thermo Fisher Scientific, USA) in $200~\mu\mathrm{L}$ of $1\times$ PBS.

Spermine tetrahydrochloride (purity $\geq 99.5\%$) was purchased from Sigma-Aldrich. Solutions of varying spermine concentrations were prepared in 20~mM HEPES containing 50~mM KCl and used for the measurements.

\subsection{Flow cell preparation and DNA attachment}

Flow cells were prepared as described previously~\cite{stor26,stor25} by sandwiching a parafilm layer, cut to form a 25~$\mu$L channel, between two thoroughly cleaned microscope coverslips (24 $\times$ 60 mm; Carl Roth, Germany), then sealing the assembly by heating on a hot plate at 80$^\circ$C for approximately 20 s. The bottom coverslip was functionalized with (3-glycidoxypropyl)trimethoxysilane (Fisher Scientific, USA) and incubated with 100~$\mu$L of 1~$\mu$m polystyrene beads (Polysciences, USA) in ethanol. These beads were used as reference markers for drift correction in the measurements. To enable liquid exchange in the flow cell, the top coverslip was prepared with 1 mm inlet and outlet holes. Next, 100~$\mu$L of 100~$\mu$g/mL anti-digoxigenin antibody (Abcam, UK) in 1$\times$ PBS was introduced to the flow cell assembly and incubated for 2 h. This was followed by passivation using 100~$\mu$L of 250 mg/mL bovine serum albumin (Carl Roth, Germany) for 2 h to minimize nonspecific interactions. The bead-coupled DNA constructs were subsequently introduced into the flow cell chamber and immobilized on the bottom coverslip through non-covalent interactions between digoxigenin and anti-digoxigenin during a 30 min incubation. Finally, the flow cell was rinsed with 1$\times$ PBS to flush out any unbound or non-specifically attached beads, and the measurement buffer containing 20 mM HEPES, 50 mM KCl, and variable concentrations of spermine was introduced into the flow chamber.

\subsection{Data acquisition}

Measurements were performed at room temperature using a custom-built single-molecule magnetic tweezers (MT) setup~\cite{stor26,walk18} [Fig.~\ref{fig:SI_exp_timetraces}(a)]. The magnetic field was generated by two permanent magnets (5 $\times$ 5 $\times$ 5~mm$^3$) positioned vertically with a 1~mm gap between them~\cite{lipf09a}. Prior to data acquisition, we applied controlled forces and torques to screen out DNA-tethered beads bound by multiple tethers or nicked tethers, using the well-known rotation-extension response of DNA~\cite{stri96,krie17a}.

The MT setup enables the application of precisely calibrated stretching forces and controlled rotation of the magnetic bead, thereby allowing direct manipulation of the linking number of the tethered molecules. To determine the force-extension response, we systematically measured tether extension with the magnets positioned at defined heights corresponding to calibrated forces ranging from 8.6 to 0.01~pN. Forces were calibrated from the transverse fluctuations of the beads with the Hadamard variance method~\cite{prit24}, using the same 20.6~kbp DNA tethers. After initial force calibration, extension data were acquired for 30 s at each force [Fig.~\ref{fig:SI_exp_timetraces}(b)], and the mean extension was calculated. We then carried out rotation-extension measurements on torsionally constrained tethers at several forces (5, 4, 3, 2.5, 2, 1.5, 1, and 0.5~pN), tracking extension as a function of the applied magnet turns. At each applied force, the linking number was varied in increments of 5 turns. At each linking number, extension data were acquired for 10~s at a camera acquisition rate of 58~Hz, from which the mean extension value was calculated.

To characterize the concentration-dependent effects of spermine on DNA rotation-extension and force-extension behavior, we performed these measurements at increasing spermine concentrations, spanning the nM, $\mu$M, and mM ranges. At each step, at least 200~$\mu$L ($>8$ cell volumes) of spermine solution at the respective concentration was flushed into the flow cell chamber and allowed to equilibrate for 10~min before data acquisition. For the DNA compaction, spermine was flushed while DNA was held at high force (8.6~pN), ensuring the DNA remained fully extended. The force was then gradually reduced to monitor the compaction process. Keeping the magnets close to the flow cell surface during flushing also prevented flow-induced rotation of the tethered beads, thereby preserving a constant linking number throughout the measurement.

\subsection{Supplementary Data}

We investigated DNA compaction in the presence of the natural tetravalent polyamine spermine. We characterized both the force-extension and rotation-extension behavior of individual dsDNA tethers across a range of spermine concentrations.

For torsionally relaxed DNA held under constant force in the presence of spermine, we found that initially, at high forces, the DNA extension is very close to that observed in the absence of spermine [Fig.~\ref{fig:SI_exp_timetraces}(b)]. Upon systematically lowering the force, we observed sudden decreases in extension and ultimately a complete collapse of the DNA tether to zero extension [Fig.~\ref{fig:fig3}(b) and Fig.~\ref{fig:SI_exp_timetraces}], consistent with spermine-induced compaction. The critical force for compaction ($f_c$) increased with spermine concentration, rising from $\approx 0.05$~pN at 50~nM to $\approx 1$~pN at 10~$\mu$M, and reaching $\approx 2$~pN at 1~mM spermine, each determined to within the spacing of the applied force steps [Fig.~\ref{fig:fig3}(b)]. While extension-time traces exhibited bead-to-bead variation in compaction kinetics, the mean compaction force remained consistent between individual tethers at a specific concentration of spermine. We note that although spermine-induced compaction was reversible, decompaction occurred only at forces well above $f_c$ under our experimental conditions. This hysteresis between compaction and decompaction is consistent with previous reports~\cite{besteman07}.

Next, to find out whether the degree of supercoiling can tune the size of plectoneme droplets in the presence of spermine, we measured DNA extension as a function of applied turns at different stretching forces. In the absence of spermine [squares in Fig.~\ref{fig:fig3}(c)], our measurements recapitulate the well-known extension-rotation curves for DNA~\cite{stri96,krie17a,vand22}: DNA extension remained largely unchanged during the initial stage of positive supercoiling, until the molecule reached a critical buckling threshold. Beyond this point, the extension decreased linearly with the number of applied turns, reflecting the formation and growth of the plectonemic phase.

Given that the critical force for compaction is concentration-dependent, we chose 1 mM spermine for the rotation-extension measurements, since the relatively higher compaction force at this concentration ($\sim 2$ pN) allowed us to readily access forces around the compaction threshold while remaining within an experimentally practical range. We performed rotation-extension measurements across a range of stretching forces. Nonetheless, we focused our analysis on the force regime at and above the compaction force (i.e. $\geq 2$ pN) [Fig.~\ref{fig:fig3}(c)], where spermine-induced sudden collapse is not expected to occur, thereby isolating the effect of spermine on plectonemes.

The extension--rotation curves in the presence of 1 mM spermine [circles in Fig.~\ref{fig:fig3}(c)] show that the buckling point is shifted to a lower number of turns compared to bare DNA. In addition, the magnitude of the slope of turns vs.\ extension in the plectonemic regime is reduced compared to bare DNA at the highest force. Both effects are due to the positive charge introduced by spermine and have been observed previously for spermine~\cite{shao12} and other ions as well~\cite{brutzer10}. Notably, these effects are not included in our computational model and are consequently not observed in the simulations. While these details preclude a direct quantitative comparison between simulations and experiments, the overall qualitative comparison is not affected by them.

We observed discrete steps in the rotation--extension behavior at 1 mM spermine. This observation suggests that multiple plectoneme domains may nucleate independently along the DNA, each locally stabilized by spermine condensation. These domains subsequently coarsen and merge into a single topological droplet as they grow sufficiently large. This progressive coalescence of plectoneme-associated condensates may result in the overall steeper post-buckling slope at lower forces, as each merging event directs more writhe into the single growing droplet.

\section{Coarsening dynamics}
\label{sec:app_coarsening}
\begin{figure}[t]
\centering
\includegraphics[width=8.6cm]{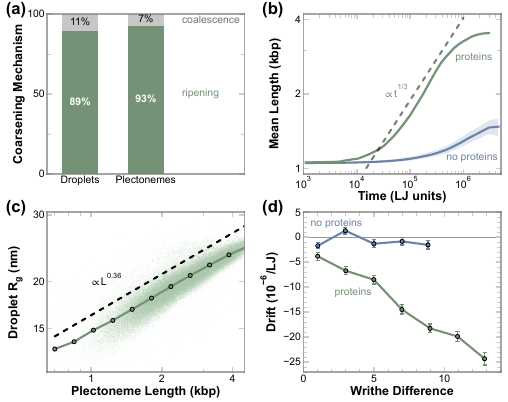}
\caption{
\textbf{Coarsening mechanism.}
(a)~Fraction of subordinate droplets and of the plectonemes present at protein addition that disappear by ripening (green) or coalescence (gray), from $788$ simulations.
(b)~Mean plectoneme length versus time after protein addition with (green) and without proteins (blue; time origins as in Fig.~\ref{fig:fig4}(b)). Dashed: $t^{1/3}$; shading: $95\%$ confidence interval.
(c)~Radius of gyration of the surviving droplet versus the length of its plectoneme for $2\times10^{4}\le t\le10^{6}$ LJ units. Points: individual snapshots; circles: binned mean; dashed: fit $R_g\propto\lplec^{0.36}$.
(d)~Drift of the writhe of the smaller of two plectonemes versus their writhe difference, with (green) and without proteins (blue).
}
\label{fig:SI_coarsening}
\end{figure}

\subsection{Simulation protocol}

\textit{Protein-free trajectories.}
We simulated $100$ independent protein-free trajectories of the $8$\,kbp tether (Sec.~\ref{sec:app_md}) at $\sig=0.04$ and $f=1$\,pN.
After equilibration, each trajectory was continued for $5\times10^{6}$ LJ units, and snapshots were stored every $10^{3}$ LJ units. In every snapshot, plectonemes were identified as described in Sec.~\ref{sec:app_dropfinder}.

\textit{Initial configurations.}
As starting points for the simulations with proteins, we selected snapshots with at least two well-separated plectonemes. After merging plectoneme segments at most $50$\,bp apart, a snapshot qualified if it contained at least two plectonemes with writhe $\wr\geq3$, two consecutive ones of which were separated by at least $300$\,bp of extended DNA, and if its linking number deviated by less than one turn from the median of all snapshots. In total, $234{,}567$ snapshots met these criteria. To obtain decorrelated configurations, we placed on each trajectory a grid with random phase and a spacing of $4\times10^{5}$ LJ units, longer than the correlation time of the plectoneme configuration ($\approx3\times10^{5}$ LJ units), and drew $1000$ snapshots uniformly from the qualifying grid points of all trajectories. Of these, $910$ contained two, $89$ three and one four plectonemes with $\wr\geq3$.

\textit{Simulations with proteins.}
Proteins were added to each initial configuration at $t=0$ ($\varphi=0.001$, at least $2000$ proteins, $\varepsilon_{\mathrm{dp}}=0.8$, $\varepsilon_{\mathrm{pp}}=1.5$; Sec.~\ref{sec:app_md}). Simulations were run in segments of $10^{5}$ LJ units. After each segment, the last snapshot was analyzed, and a simulation was stopped once two consecutive checks found a single droplet, i.e., a single protein cluster of at least $10$ proteins (single-linkage clustering with cutoff $3\sigma$), and a single plectoneme with $\wr\geq2$; otherwise, it was stopped after at most $2\times10^{7}$ LJ units. We analyzed $788$ simulations, all of which reached the stopping criterion.

\subsection{Analysis}

Snapshots were analyzed every $10^{3}$ LJ units. Plectonemes were counted after merging segments at most $50$\,bp apart and grouping segments at most $200$\,bp apart into one plectoneme; only plectonemes with $\wr\geq2$ were counted. The same rules were applied to all simulations without proteins, including the supercoiling scans of Fig.~\ref{fig:fig4}(c). Droplets are protein clusters of at least $10$ proteins, as above. Droplets and plectonemes were followed in time via the overlap of their DNA contacts and contour positions, which shows how each subordinate droplet disappears: by dissolving while larger droplets grow (ripening) or by merging with another droplet (coalescence).

In Fig.~\ref{fig:fig4}(b), the curves with proteins are averages over the $788$ simulations; once a simulation has stopped, its final state is carried forward. The protein-free curve uses all $234{,}567$ qualifying snapshots of the protein-free trajectories as time origins and averages the number of plectonemes in the subsequent snapshots, i.e., it shows the relaxation of protein-free tethers from the same kind of initial state. It reaches the protein-free equilibrium of $\approx1.7$ plectonemes after $\approx3\times10^{6}$ LJ units. Shaded bands are $95\%$ confidence intervals from bootstrapping over independent trajectories. The triangles mark the median time until the initial pair has lost a plectoneme, i.e., until a single plectoneme is observed in at least $5$ consecutive snapshots (Kaplan--Meier estimate; protein-free trajectories are censored at their end): $2.3\times10^{5}$ LJ units with and $1.1\times10^{6}$ LJ units without proteins.

In Fig.~\ref{fig:fig4}(c), the protein-free supercoiling scans of Fig.~\ref{fig:fig2} were recounted with the rules above. Error bars are standard errors between independent replicas (for the few points with a single replica, the autocorrelation-corrected standard error). With proteins, a single droplet forms beyond the onset at all forces. At $f=1$\,pN and $\sig=0.035$--$0.05$, the mean number of plectonemes with proteins is $1.25$--$1.34$ although a single droplet is present: small, droplet-free plectonemes occasionally nucleate next to the droplet.

In Fig.~\ref{fig:fig4}(d), $R_g$ is that of the droplet that remains at the end of the simulation. The exponent was obtained from a power-law fit to the logarithmically binned mean for $2\times10^{4}\le t\le10^{6}$ LJ units, $R_g\propto t^{0.095\pm0.002}$ (bootstrap over simulations).

\subsection{Coarsening mechanism}

Ripening dominates the coarsening [Fig.~\ref{fig:SI_coarsening}(a)]. Of $797$ subordinate droplets, $713$ ($89\%$) dissolve while a larger droplet grows, and $84$ ($11\%$) merge with another droplet. Dissolving droplets shrink to a median of $4$ proteins before they disappear, whereas merging droplets still contain $16$. Of the plectonemes present at protein addition, $789$ vanish and $61$ merge ($93\%$ ripening); $43$ had not yet disappeared when the simulation stopped.

With proteins, the mean plectoneme length grows from $1.06$ to $3.5$\,kbp [Fig.~\ref{fig:SI_coarsening}(b)], at intermediate times roughly as $t^{1/3}$, as for one-dimensional coarsening of plectonemes~\cite{forte2019}. Part of this growth stems from the conversion of twist into writhe: with proteins, the total plectonemic length increases from $2.3$ to $3.6$\,kbp. Without proteins, the mean length only relaxes from $1.05$\,kbp toward its equilibrium value of $\approx1.5$\,kbp.

The droplet size follows the length of its plectoneme, $R_g\propto\lplec^{0.362\pm0.002}$ [Fig.~\ref{fig:SI_coarsening}(c); $4.2\times10^{5}$ snapshots of $788$ simulations, bootstrap over simulations], close to the $\lplec^{1/3}$ expected if the droplet volume is proportional to the plectoneme length~\cite{sommer22pac}. Together with a plectoneme length growing as $t^{1/3}$, this gives $R_g\propto t^{1/9}$, close to the measured $t^{0.10}$.

Finally, we characterized the exchange of writhe between two plectonemes [Fig.~\ref{fig:SI_coarsening}(d)]. In snapshots with exactly two plectonemes, we followed the writhe $\wr_s$ of the smaller one over $1$--$10$ snapshots. To remove changes of the total writhe, we used its share of the writhe of the pair. The drift $v$ is the slope of the mean change of $\wr_s$ versus time, binned in the writhe difference $\wr_l-\wr_s$ (bins containing at least $5\%$ of the snapshots). Without proteins, $v$ is close to zero ($|v|\lesssim2\times10^{-6}$ turns per LJ unit), i.e., writhe performs an unbiased random walk between the plectonemes. With proteins, writhe drifts toward the larger droplet, with $|v|$ increasing from $4\times10^{-6}$ to $24\times10^{-6}$ turns per LJ unit with the writhe difference: writhe has a lower free energy in the larger droplet. Because $v$ is measured on the detected plectonemes, only the comparison between the two cases is meaningful.

\bibliography{references}